\documentclass[11pt,reqno]{amsart}
\usepackage{geometry}
\usepackage{amssymb, amsmath,amsthm, amsfonts}
\usepackage{centernot}
\usepackage[colorlinks=true]{hyperref}
 \hypersetup{
     colorlinks=true,
     linkcolor=blue,
     filecolor=blue,
     citecolor = RoyalBlue,      
     urlcolor=cyan,
     }
\usepackage[dvipsnames]{xcolor}
\usepackage{url}
\usepackage{graphicx}
\usepackage{enumerate}
\usepackage{epstopdf}
\usepackage{subfig}
\usepackage{mathtools}
\usepackage{csquotes}
\usepackage{ifthen}
\usepackage{tikz}
\usepackage{natbib}
\usepackage{cite}
\usepackage{verbatim}
\usepackage{pifont}
\usepackage{bm}  
\usepackage{stackengine}
\usepackage{bbm}
\allowdisplaybreaks
\usepackage{accents}
\patchcmd{\subsubsection}{\itshape}{\bfseries}{}{}

\usepackage{xpatch}
\makeatletter
\AtBeginDocument{\xpatchcmd{\@thm}{\thm@headpunct{.}}{\thm@headpunct{}}{}{}}
\makeatother
\makeatletter
\let\save@mathaccent\mathaccent
\newcommand*\if@single[3]{%
  \setbox0\hbox{${\mathaccent"0362{#1}}^H$}%
  \setbox2\hbox{${\mathaccent"0362{\kern0pt#1}}^H$}%
  \ifdim\ht0=\ht2 #3\else #2\fi
  }
\newcommand*\rel@kern[1]{\kern#1\dimexpr\macc@kerna}
\newcommand*\widebar[1]{\@ifnextchar^{{\wide@bar{#1}{0}}}{\wide@bar{#1}{1}}}
\newcommand*\wide@bar[2]{\if@single{#1}{\wide@bar@{#1}{#2}{1}}{\wide@bar@{#1}{#2}{2}}}
\newcommand*\wide@bar@[3]{%
  \begingroup
  \def\mathaccent##1##2{%
    \let\mathaccent\save@mathaccent
    \if#32 \let\macc@nucleus\first@char \fi
    \setbox\z@\hbox{$\macc@style{\macc@nucleus}_{}$}%
    \setbox\tw@\hbox{$\macc@style{\macc@nucleus}{}_{}$}%
    \dimen@\wd\tw@
    \advance\dimen@-\wd\z@
    \divide\dimen@ 3
    \@tempdima\wd\tw@
    \advance\@tempdima-\scriptspace
    \divide\@tempdima 10
    \advance\dimen@-\@tempdima
    \ifdim\dimen@>\z@ \dimen@0pt\fi
    \rel@kern{0.6}\kern-\dimen@
    \if#31
      \overline{\rel@kern{-0.6}\kern\dimen@\macc@nucleus\rel@kern{0.4}\kern\dimen@}%
      \advance\dimen@0.4\dimexpr\macc@kerna
      \let\final@kern#2%
      \ifdim\dimen@<\z@ \let\final@kern1\fi
      \if\final@kern1 \kern-\dimen@\fi
    \else
      \overline{\rel@kern{-0.6}\kern\dimen@#1}%
    \fi
  }%
  \macc@depth\@ne
  \let\math@bgroup\@empty \let\math@egroup\macc@set@skewchar
  \mathsurround\z@ \frozen@everymath{\mathgroup\macc@group\relax}%
  \macc@set@skewchar\relax
  \let\mathaccentV\macc@nested@a
  \if#31
    \macc@nested@a\relax111{#1}%
  \else
    \def\gobble@till@marker##1\endmarker{}%
    \futurelet\first@char\gobble@till@marker#1\endmarker
    \ifcat\noexpand\first@char A\else
      \def\first@char{}%
    \fi
    \macc@nested@a\relax111{\first@char}%
  \fi
  \endgroup
}
\makeatother
\DeclareMathOperator*{\argmax}{arg\,max}
\DeclareMathOperator*{\argmin}{arg\,min}
\DeclareSymbolFont{bbold}{U}{bbold}{m}{n}
\DeclareSymbolFontAlphabet{\mathbbold}{bbold}

\newtheorem{theorem}{Theorem}

\newtheorem{prop}{Proposition}

\newtheorem{remark}{Remark}

\newcommand{\norm}[1]{\left\Vert#1\right\Vert}
\newcommand{\abs}[1]{\left\vert#1\right\vert}

\newcommand{\cA}{\mathcal{A}}
\newcommand{\cB}{\mathcal{B}}
\newcommand{\cC}{\mathcal{C}}
\newcommand{\cD}{\mathcal{D}}
\newcommand{\cE}{\mathcal{E}}

\newcommand{\cG}{\mathcal{G}}
\newcommand{\cH}{\mathcal{H}}

\newcommand{\cJ}{\mathcal{J}}

\newcommand{\cL}{\mathcal{L}}
\newcommand{\cM}{\mathcal{M}}

\newcommand{\cP}{\mathcal{P}}
\newcommand{\cQ}{\mathcal{Q}}

\newcommand{\cS}{\mathcal{S}}

\newcommand{\cU}{\mathcal{U}}
\newcommand{\cV}{\mathcal{V}}
\newcommand{\cW}{\mathcal{W}}
\newcommand{\cX}{\mathcal{X}}

\newcommand{\cZ}{\mathcal{Z}}

\newcommand{\CC}{\mathbb{C}}

\newcommand{\EE}{\mathbb{E}}

\newcommand{\HH}{\mathbb{H}}

\newcommand{\NN}{\mathbb{N}}

\newcommand{\PP}{\mathbb{P}}

\newcommand{\RR}{\mathbb{R}}

\newcommand{\UU}{\mathbb{U}}

\newcommand{\supp}{\mathrm{supp}}

\newcommand{\tr}[1]{\operatorname{Tr}{\left[#1\right]}}
\newcommand{\kl}[2]{\mathsf{D}_\mathsf{KL}\left(#1\middle\|#2\right)}

\definecolor{cblue}{rgb}{0.16, 0.32, 0.75}

\newcommand{\nll}{\centernot{\ll}}

\def\h2{\tilde h}

\def\hm1{\hat h_{-1}}

\allowdisplaybreaks

\newcommand{\qrel}[2]{\mathsf{D}\left(#1\middle\|#2\right)}
\newcommand{\mrel}[3]{\mathsf{D}_{\mathsf{#3}}\left(#1\middle\|#2\right)}

\newcommand{\bra}[1]{\langle#1|}
\newcommand{\ket}[1]{|#1\rangle}
\newcommand{\ketbra}[2]{|#1\rangle \langle #2|}

\newcommand{\innp}[2]{\langle #1,#2\rangle}

\usepackage{xcolor}

\begin{document}
\title[Prediction via Hilbert Space Embeddings]{Quantum Maximum Likelihood Prediction  via Hilbert Space Embeddings}
\author[S. Sreekumar]{Sreejith Sreekumar}
\address[S. Sreekumar]{L2S, CNRS, CentraleSupélec,  University of Paris-Saclay, France}
\email{sreejith.sreekumar@centralesupelec.fr}

\author[N. Weinberger]{Nir Weinberger}
\address[N. Weinberger]{
Electrical and Computer Engineering, Technion,  Israel
}
\email{nirwein@technion.ac.il}
\begin{abstract}
Maximum likelihood prediction (MLP) is a core task at the heart of modern large language models.   
Here,  we study a quantum version of this task for a simplified data model consisting of independent and identically distributed  samples, as a first step. The quantum maximum likelihood predictor (QMLP) is obtained by embedding of empirical probability distributions into quantum states  and performing a minimization of quantum relative entropy over a given class of states. 
We derive non-asymptotic performance guarantees for QMLP in terms of convergence rates and concentration inequalities, both in trace norm and quantum relative entropy.   Our approach provides a unified framework to handle MLP within both classical and quantum LLMs. We also consider the related problem of quantum information projection and generalize the quantum Pythagorean theorem to mixture families specified by possibly non-self-adjoint linear constraints. We further show that the Pythagorean inequality continues to hold in the infinite-dimensional setting whenever the convex information-projection problem attains a finite minimum.    
\end{abstract}
\keywords{Quantum maximum likelihood prediction, large language models, embeddings, quantum Pythagorean theorem, quantum information projection}
\maketitle

\section{Introduction}
Prediction is a primitive task with numerous applications in machine
learning, statistics, information theory, physics, finance, and engineering.
The basic prediction problem pertains to providing an informed guess
about the next outcome, $X_{n+1}$, of a sequence, given the previous
observations $X_{1},\ldots,X_{n}$. This problem has been studied from multiple
perspectives, e.g.,  machine learning  \citep{cesa2006prediction}
and information theory \citep{merhav2002universal}.  In the classical theory, the performance of the predictor
degrades with the dimension of the parameter vector used to determine
the predictor. Specifically, under the common log loss (cross-entropy), the expected regret in predicting that a sample from a probability distribution $P$ follows
a probability distribution $Q$ is given by the Kullback--Leibler (KL) divergence, $\kl{P}{Q}$. In the minimax framework, the regret
scale linearly with this dimension of the model class of $Q$ (e.g., \citet{Krichevsk1981performance, rissanen2003universal,yang1999information,haussler1997mutual}, \citet[Sec. 13.4]{polyanskiy-wu-ITbook}). 
However, in modern applications,  the parameter dimensionality could be large, e.g.,  $(i)$ the number of parameters in the architecture  of a deep neural network, or any of the various refinements of their effective dimension (e.g. \citet{golowich2020size}), or $(ii)$ the cardinality of the vocabulary of token sequences in large language models (LLMs). Despite this seemingly large effective dimension, modern prediction algorithms work remarkably well in practice,  suggesting that training and prediction must rely on some inductive bias, whether explicit or implicit.

Here, we assume that the inductive bias is captured by embedding  the letters $x\in\cX$ of the alphabet into  vectors in a finite-dimensional Hilbert space of dimension $d$. Similarly to the classical prediction problem, a sequence of $n$ independent and identically distributed (i.i.d.) symbols
$X_{1}^{n}=(X_{1},X_{2},\ldots,X_{n})$ from an unknown distribution
$P$ is observed, and the model should learn to predict the next symbol $X_{n+1}$
via a probability distribution $Q$, while aiming to minimize the
mean log loss $-\EE_P[\log Q(X_{n+1})]$.   Intuitively, since the embedding
is to a finite-dimensional Hilbert space of dimension much smaller than the vocabulary, the curse of large vocabulary is alleviated, and the discrete geometry between words in the vocabulary (Hamming distance or ``zero-one'' similarity) is replaced by the
inner product in the embedded space. The embedding thus represents the inductive bias necessary for efficient learning. 

A prominent representative of this idea are LLMs, in the sense that during inference they induce a probability distribution over the next token (symbol, word, sentence, etc.) based on previous ones.  The internal representation of the LLM (say key, query and value matrices, see Appendix \ref{Sec:LLMs}) can be thought of as an embedding to a Hilbert space, where the embedding is learned via extensive training and is sufficiently flexible to encapsulate common dependencies within the data model. Such a viewpoint provides a geometry in which token similarity is quantified by inner products,  whereas the original discrete space lacks an intrinsic geometric structure. Additionally, this Hilbert space equipped with the notion of orthogonality provides a natural setting where  optimization algorithms such as stochastic gradient descent (SGD) can be performed efficiently and analyzed.

Our proposed predictor utilizes the learned embedding by a mapping
of the empirical distribution of the symbols $X_{1}^{n}$ into the
embedded domain. A natural baseline is the \emph{mean embedding}, in which a probability distribution $Q$
is mapped to $\EE_{X\sim Q}[\ket{\varphi(X)}]$, where  the notation $\ket{\cdot}$ (called ket)  borrowed from quantum theory represents the usual vector in a Hilbert space. Although widely used \citep{muandet2017kernel}, this embedding induces a Euclidean geometry on the space of probability distributions, which does not align with the geometry of the probability simplex and  information-theoretic objectives. As recently argued and explored by \citet{Bach-2023}, a
more natural embedding of probability distributions for learning purposes is to the \emph{covariance
operator} $\EE_{X\sim Q}[\ket{\varphi(X)}\bra{\varphi(X)}]$, or,
in quantum-information terms \citep{wilde2017quantum,Hayashi-book-2016}, to a \emph{density operator}. The resulting geometry enables a formulation of MLP formulation based on minimization of
quantum relative entropy, which is a natural generalization of the KL divergence (which is the expected log loss up to an entropy correction). We refer to this simple predictor
as  \emph{quantum maximum likelihood predictor
}(QMLP). The desired classical predictor is  obtained by processing the QMLP to an output probability distribution (in the case of a quantum LLM this involves applying a measurement, i.e., a positive operator-valued measure (POVM),  according to the Born rule). This induces the bias that closeness in quantum relative entropy correlates with predictive accuracy on the output alphabet.

Given the above context for our proposed predictor, several natural basic questions emerge: What are the
structural and statistical properties of
performing MLP in the space of embedded quantum 
states rather than the original space of probability distributions? 
As we show below, the smaller dimension of the Hilbert space (e.g., in the context of LLMs), compared to the original alphabet size of the discrete distributions (which can be large or even infinite), enables QMLP to have a much better convergence rate than its analogue in the distribution space. 
\subsection{Contributions}

Our main contributions can be summarized as follows:
\begin{enumerate}[(i)]
\item We  show that  under model-class symmetries  (unitary invariance and closure under pinching operation), the QMLP objective reduces to a classical KL divergence objective on eigenvalues (Proposition \ref{prop:QMLPrel}). However, these symmetry assumptions are hardly realized in practice, in which case the general QMLP  is still useful.
    \item 
    We obtain non-asymptotic upper bounds (Theorem \ref{Thm:convergence-MLpred}) quantifying the expected rate of convergence of QMLP (and MLP) to the target state  both in trace norm and quantum relative entropy. We also show the corresponding concentration inequalities. These bounds depend on the dimension $d$ of the Hilbert space, the sample size $n$, and the minimal positive eigenvalue of the embedded quantum state corresponding to the true data distribution. For instance, our bounds show that the rate of convergence of the expected squared trace-norm error is  $\tilde O(d/\sqrt{n})$ (up to an additive approximation slack factor) in general and $\tilde O(d^3/n)$ when the class of quantum models is sufficiently expressive, where $\tilde O(\cdot)$ hides logarithmic factors and dependence on the minimal eigenvalue. Under additional conditions on the expressivity of quantum models measured in terms of quantum relative entropy, the corresponding rates of convergence under quantum relative entropy are $\tilde O(d/\sqrt{n})$  and $O\big(\big(\norm{\rho_p^{-1}}+\log d\big)/n\big)$, respectively. We also derive upper bounds (Proposition \ref{Prop:covnumdensop}) on the covering number of quantum states under relative entropy. 
    \item Of independent interest,  we consider the related problem of  information projection in quantum information geometry \citep{Amari-Nagaoka-2000,Hayashi-book-2016}. We generalize the finite-dimensional Pythagorean theorem (Theorem \ref{Thm:quantPyth}) to mixture families described by possibly non-self-adjoint constraints. We further extend the Pythagorean inequality (Theorem \ref{Thm:Qpyth-infdim}) to an infinite-dimensional setting under the sole requirement that the convex information-projection problem attain a finite minimum. 
\end{enumerate}
\subsection{Related work}
 Embedding probability distributions into Hilbert spaces for inference has been explored in a variety of other contexts. For instance, the mean embedding \citep{BerlinetThomasAgnan2004,muandet2017kernel} and the associated maximum mean discrepancy (MMD) metric has been used for designing kernel based two-sample tests \citep{GBRSS-2006}, defining independence/dependence measures \citep{FukumizuBachJordan2004},  feature extraction \citep{Smola-2007} and graphical models \citep{Song-Gretton-Guestrin-2010} among other applications. The covariance embedding \citep{Bach-2023} of probability distributions into density operators has also found several applications. To mention a few, \citet{Hoyos-2024} considered a representation Jensen-Shannon divergence defined in terms of von Neumann entropy of  embedded density operators, \citet{SantoroPanaretos2025} proposed a kernel-based likelihood test utilizing both the mean and covariance embedding, and \citet{kachaiev2024learning}
proposed unsupervised learning of the kernel via maximum von Neumann entropy. The covariance embedding of probability distributions and the associated QMLP also captures at a high-level the problem of prediction in quantum LLMs \citep{basile-tamburini-QLM}, where the expressivity of quantum architecture used for prediction is reflected in the class of models over which QMLP is optimized. Another related topic corresponds to quantum generative modeling and simulation of probability distributions, in which the hope is to use quantum models to efficiently simulate probability distributions which are difficult to model classically.   Two popular frameworks  used in this regard pertains to quantum circuit Born machines \citep{Liu-Wang-2018,Benedetti2019generative} and quantum Boltzmann machines \citep{Benedetti-2017-QBMs,Kieferova-Wiebe-2017,Amin-QBM-2018}, which uses quantum circuits and thermal (or Gibbs) states, respectively, to achieve the modeling objective.
\section{Preliminaries}
\subsection{Notation} \label{Sec:notation}
Let $\cX$ be a discrete set or a subset of $\RR^d$. Denote the set of all Borel probability measures whose support is contained in $\cX$ by $\cP(\cX)$.  Let $\mu$ denote a sigma-finite measure on $\cX$.   For $P \ll \mu$ (i.e., $P$ absolutely continuous with respect to $\mu$), let $p$ denote its Radon-Nikodym derivative with respect to $\mu$. Unless specified otherwise, we will take $\mu$ to be the counting measure when $\cX$ is discrete or Lebesgue measure when $\cX \subseteq \RR^d$, in which case $p$ becomes the usual probability mass function (pmf) or probability density function (pdf). 

Let $\HH$ denote a complex (or real) separable Hilbert space. Unless specified otherwise, we will assume that $\HH$ is complex. When $\HH$ is of dimension $d$, we will denote it as $\HH_d$.  For an inner product $\innp{\cdot}{\cdot}_{\HH}$ on $\HH$, we will use the physics inspired convention in which the inner product is anti-linear (or conjugate) in the first argument and linear in the second. Whenever there is no confusion, we will denote $\innp{\cdot}{\cdot}_{\HH}$ simply by $\innp{\cdot}{\cdot}$.   The space of continuous linear operators on $\HH$ is denoted by $\cL(\HH)$, equipped with operator norm topology.  The set of density operators or quantum states is denoted as $\cS(\HH)$, i.e., a self-adjoint operator $\rho$ such that $\rho \geq 0$ and $\tr{\rho}=1$. The operator (or Schatten-$\infty$) and trace  norms are denoted by $\norm{\,\cdot\,}$ ($=\norm{\,\cdot\,}_{\infty}$) and  $\norm{\,\cdot\,}_1$,  respectively.

We will employ the bra-ket notation from quantum information theory (see e.g. \citet{Hayashi-book-2016}). To mention it briefly, let $h$ and $\tilde h$ be any two elements of some Hilbert space $\HH$. Then, $\ket{h}$, $\bra{\tilde h}$ and $\ketbra{h}{\tilde h}$  corresponds, respectively, to an element $h \in \HH$, a linear functional on $\HH$ with action $\bra{\tilde h} \ket{h}=\innp{\tilde h}{h}$, and a linear operator with action $\ketbra{h}{\tilde h} \ket{\bar h}=\innp{\tilde h}{\bar h} \ket{h}$ for every $\bar h \in \HH$, respectively. 
For $L \in \cL(\HH)$, $\lambda_L$ denotes the vector composed  of eigenvalues of $L$. When $\rho \in \cS(\HH)$,  $\lambda_{\rho}$ is a pmf. The support of $L \in \cL(\HH)$, which is defined as the orthogonal complement of kernel of $L$ in $\HH$, is denoted by $\supp(L)$. $L_1 \ll L_2$ denotes that $\supp(L_1) \subseteq \supp(L_2)$ and $L_1 \nll L_2$ means that $\supp(L_1) \not\subseteq \supp(L_2)$. The support of $\Sigma \subseteq \cL(\HH)$ is the closed subspace $\supp(\Sigma)\coloneqq \overline{\operatorname{span}(\cup_{\rho \in \Sigma}\supp(\rho))}^{\,\norm{\cdot}_{\HH}}$. 
For $a,b \in \RR$, $a \vee b$ and $a \wedge b$ denote $\max\{a,b\}$ and $\min\{a,b\}$, respectively. 

Next, we introduce the preliminary notions used in the paper.
\subsection{Quantum Relative Entropy and Measured Relative Entropy}
The \emph{quantum relative entropy} \citep{U62} between $\rho,\sigma \in \cS(\HH)$ is \\[-5 pt]
\begin{align}
    \qrel{\rho}{\sigma}\coloneqq \begin{cases}
      \tr{\rho(\log \rho-\log \sigma)}, &\mbox{ if } \rho \ll \sigma,\\
      \infty, &\mbox{otherwise},
    \end{cases} \label{eq:quant-rel-ent}
\end{align}
whenever the trace expression is well defined. Otherwise, $\qrel{\rho}{\sigma}$ denotes the standard extended lower-semicontinuous Umegaki relative entropy, defined via  extension of the finite-dimensional formula. 
The quantum relative entropy admits the variational expressions \citep{Petz1988,Berta2015OnEntropies}:
\begin{align}
    \qrel{\rho}{\sigma}=\sup_{H} \tr{H\rho}-\log \tr{e^{H+\log \sigma}} =\sup_{H} \tr{H\rho}- \tr{e^{H+\log \sigma}}+1, \label{eq:varexpqrelent} 
\end{align}
where the supremum is taken over all bounded self-adjoint operators $H$. 
The \emph{measured relative entropy} \citep{Donald1986,P09} between $\rho,\sigma \in \cS(\HH)$ is defined as
\begin{equation}
\mrel{\rho}{\sigma}{\mathsf{M}} \coloneqq\sup_{\cZ,\left\{  M_z\right\}  _{z\in\cZ}}
\sum_{z\in\cZ}\tr{M_z\rho}\log \!\left(  \frac{\tr{M_z\rho}}{\tr{M_z\sigma}}\right),
\label{eq:measured-rel-def}
\end{equation}
where the supremum is over all finite sets $\cZ$ and positive operator-valued measures\,\footnote{A POVM $\left\{  M_z\right\}_{z\in\cZ}$  on a separable Hilbert space $\HH$, indexed by a discrete $\cZ$, is a set of positive semidefinite operators $ M_z$ such that $\sum_{z \in \cZ} M_z$ equals the identity operator on $\HH$.} (POVMs) $\left\{  M_z\right\}_{z\in\cZ}$ indexed by~$\cZ$. 
By \emph{data-processing inequality} for quantum relative entropy, $   \qrel{\rho}{\sigma} \geq  \mrel{\rho}{\sigma}{\mathsf{M}}$. 
This shows that if two density operators $\rho$ and $\sigma$ are close in the quantum relative entropy, then the KL divergence between any two measured probability distributions is only  lower (or equal). Moreover, when $\rho,\sigma>0$, equality holds  if and only if $\rho$ and $\sigma$ commute, and then this common value also coincides with $\kl{\lambda_{\rho}}{\lambda_{\sigma}}$, where $\lambda_{\rho}$ denotes the probability mass function (pmf) composed of eigenvalues of $\rho$, and  the KL divergence between probability measures $P,Q \in \cP(\cX)$ is
\begin{align}
   \kl{P}{Q} \coloneqq \begin{cases}
      \EE_{P}\left[\log \frac{dP}{dQ}\right], &\mbox{ if } P \ll Q,\\
      \infty, &\mbox{otherwise}.
    \end{cases} 
\end{align}
The von Neumann entropy of  $\rho \in \cS(\HH)$ is $\mathsf{H}(\rho)\coloneqq -\tr{\rho \log \rho}$, which is non-negative and upper bounded by $\log d$ when $\rho \in \cS(\HH_d)$.
\section{Prediction task}
Let $m,n \in \NN$ and $\cX$ be a discrete (countable) set or a subset of $\RR^m$. Let $\cQ \subseteq \cP(\cX)$ be a class of probability measures.  We will assume that $\cQ$ is a compact (with respect to weak topology) convex set and contains a $Q>0$.
The classical prediction problem can then be framed as determining the best predictor (or model for prediction) $\hat Q_n: \cX^n \rightarrow \cQ$ based on the samples $X^n\sim P^{\otimes n}$, where $P \in \cP(\cX)$. The accuracy of prediction is measured by a loss function $\mathrm{d}:\cX \times \cP(\cX) \rightarrow [0,\infty]$. A particular loss function of interest is the log loss function $\mathrm{d}_{\mathrm{L}}(x,Q)\coloneqq -\log q(x)$, where $q$ denotes the probability mass function (p.m.f.) in the discrete case and probability density function (p.d.f.) in the continuous case. Let $\{\hat Q_n(\cdot)\}$ be the shorthand for $\{\hat Q_n(x^n)\}_{x^n\in \cX^n}$. 
Given $x^n$, we find the predictor that minimizes the cumulative (or equivalently the  average) empirical loss  according to the log loss function. Consider the case of discrete $\cX$.  Since the empirical loss can be written as
 \begin{align}
\inf_{Q \in \cQ}\frac{1}{n}\sum_{i=1}^n \mathrm{d}_{\mathrm{L}}(x_i,Q)=\mathsf{H}(\hat P_{x^n})+\inf_{Q \in \cQ} \kl{\hat P_{x^n}}{Q},
 \end{align}
where $\mathsf{H}(P)$ denotes the Shannon entropy of $P$, this procedure is equivalent to finding the predictor\begin{align}
   \hat Q_{n}^{\star}(x^n)\coloneqq \argmin_{Q \in \cQ} \kl{\hat P_{x^n}}{Q} =\argmax_{Q \in \mathcal{Q}} \prod_{i=1}^n q(x_i), \label{eq:mlest-emp}
\end{align}
when the minimum (and maximum) above exists. We refer to $\hat Q_{n}^{\star}(x^n)$ as an MLP when it exists. MLP when $\cX \subseteq \RR^m$ can be defined similarly using the first equality in \eqref{eq:mlest-emp}.   When multiple solutions to the optimization in \eqref{eq:mlest-emp} exist, MLP will mean any one of them chosen arbitrarily.

Consider the optimization (minimization) problem in \eqref{eq:mlest-emp}. Fixing $x^n \in \cX^n$ and 
setting  $Q=\hat Q_{n}(x^n)$, one may without loss of generality consider the following optimization problem:
\begin{align}
\inf_{Q \in \cQ} \kl{P}{Q}=\min_{Q \in \cQ} \kl{P}{Q}=\kl{P}{Q_P^{\star}}.\label{eq:MLopt} 
\end{align}
In the above, the equalities  follow by noting that the resulting infimum is achieved by some $Q_P^{\star}$ due to lower semicontinuity of $\kl{P}{Q}$ in $(P,Q)$ and compactness of $\cQ$, both with respect to weak topology \citep{vanEvren_Reyni_Div2014}. 
Since $\cQ$ is also convex, there exists a distribution $Q'$ whose support contains the support of all other $Q \in \cQ$. Define the support of $\cQ$, $\supp(\cQ)=\supp(Q')$, for such a $Q'$. When $\supp(P)=\supp(\cQ)$ and the value of the minimum in \eqref{eq:MLopt} is finite, $Q_P^{\star}$ is unique (up to  $P$ almost everywhere equivalence).

 Even when $\cX$ is a  discrete, solving \eqref{eq:mlest-emp} to find the maximum likelihood estimate quickly becomes infeasible when $|\cX|$ is large and a general $\cQ \subset \cP(\cX)$,  since it involves optimization over a subset of the  simplex of dimension $|\cX|-1$. This is a typical case in LLMs, for example, when  $\cX=\cW^{\otimes k}$ is the set of all possible sentences of length $k$ composed of tokens or words selected from a set $\cW$. Further, although \eqref{eq:mlest-emp} is a convex optimization problem, common algorithms such as SGD are not directly applicable on an arbitrary discrete set  $\cX$ without a metric structure or explicit parameterizations. Indeed, modern LLMs overcome this challenge by representing tokens or words as vectors in a Euclidean space where gradient-based optimization algorithms can be applied. 
  In the following, we interpret this through Hilbert-space embeddings, quantum information projection, and bounds on QMLP performance.
\subsection{Embeddings  and Quantum Maximum Likelihood  Predictor}\label{Sec:Emb-QMLP}
When considering an embedding of probability distributions into Hilbert spaces, a natural possibility, which has been extensively studied in the context of reproducing kernel Hilbert spaces (RKHSs), is the mean embedding $P \mapsto \EE_P[\varphi(X)]$, where $\varphi$ is the feature map corresponding to an RKHS (see Appendix \ref{Sec:RKHS}). However, as noted in \citet{Bach-2023}, the mean embedding induces an MMD metric between probability distributions in the RKHS,  which does not have a direct connection to log loss (and information theory).  Accordingly, \citet{Bach-2023}  proposed a second-order covariance embedding as a suitable framework to study interaction between probability distributions and perform sample-based inference. To define it, let  $\HH$ be a  Hilbert space and  $\varphi:\cX \rightarrow \HH$ be a Borel-measurable map such that $\norm{\varphi(x)}_{\HH}=1$ for all $x \in \cX$. For a  probability measure $P$ with density $p$ with respect to a  dominating positive $\sigma$-finite measure $\mu$ (say counting measure in the discrete case and Lebesgue measure in the continuous case), set
\begin{align}
    \rho_p= \int_{\cX} p(x) \ket{\varphi(x)} \bra{\varphi(x)}  d \mu(x). \label{eq:covembed}
\end{align}
The covariance embedding  $\phi: \cP(\cX) \rightarrow  \cL(\HH)$ is then the linear map  induced by \eqref{eq:covembed}.

Here, we explore the utility of covariance embeddings for  prediction. 
Note that $\rho_p \in \cS(\HH)$ (i.e., a quantum state) since it is positive semidefinite and has unit trace (see \eqref{eq:traceone} in Appendix \ref{Sec:RKHS}). Hence, $\phi$ is a map of $\cP(\cX)$ into $\cS(\HH)$. Also, recall that  $\cS(\HH)$  is a subset of trace-class operators on $\HH$, and hence lies in a Hilbert space within\footnote{For  $\HH_d$, the whole $\cL(\HH_d)$ becomes a  Hilbert space equipped with the Hilbert-Schmidt inner product.} $\cL(\HH)$ equipped with the Hilbert-Schmidt inner product  $\innp{\rho}{\sigma} \coloneqq \tr{\rho^*\sigma}$, where $\rho^*$ denotes the adjoint of $\rho$. Several interesting properties of the map $\phi$ when $\varphi$ is the feature map corresponding to an RKHS with kernel $\{K(x,y)\}_{(x,y)\in \cX \times \cX}$  were studied in \citet{Bach-2023}. As noted therein,  the image of $\phi$ lies within a  Hilbert space that is isomorphic to the RKHS generated by the kernel $\{|K(x,y)|^2\}_{(x,y)\in \cX \times \cX}$. Also, the quantum data-processing inequality  yields $\kl{P}{Q} \geq  \qrel{\rho_p}{\rho_q}$.

Given the embedding  of probability distributions as density operators in a Hilbert space (which we interpret as an internal representation), we model learning of the predictor  as selecting a density operator $\sigma$ from a model class $\Sigma$ determined by the QMLP. To obtain a classical distribution on the output vocabulary, we apply a measurement channel (or POVM) $\cM_n$, viewed as a quantum-to-classical channel \citep{Hayashi-book-2016,wilde2017quantum}. The appropriate embedding $\varphi$ and measurement are selected during training and then held fixed during the learning. Let 
\begin{align}
    \hat \rho_n(x^n) \coloneqq \frac{1}{n} \sum_{i=1}^n \ket{\varphi(x_i)}\bra{ \varphi(x_i)},  \label{eq:denopemb} 
\end{align}
denote the embedding of the empirical distribution $\hat p_n$ of $x^n$, i.e., $\hat \rho_n(x^n) = \rho_{\hat p_n} \coloneqq \phi(\hat p_n)$. Let  $\hat P_n$ and $\hat Q_n$ denote the measurement-output distributions corresponding to $\hat \rho_n(x^n)$ and $\sigma$. Prediction accuracy is quantified by the minimal average log loss, which is equivalent to  $ \min_{Q \in \cQ} \mathsf{D}_{\mathsf{KL}}\big(\hat P_n\|Q\big)$. Here,   $\cQ\coloneqq \{\cM_n(\sigma):\sigma \in \Sigma\}$ and $\cM_n(\sigma)$ denotes the classical  distribution obtained as the output of the measurement channel with input $\sigma$. Since the measurement $\cM_n$ is unspecified, one may consider the above optimization with measured relative entropy in place of KL divergence and replacing $\hat P_n$, $Q$ and $\cQ$ by  $\hat \rho_n(x^n), \sigma$ and $\Sigma$, respectively. 
However, the measured relative entropy is typically intractable, as it requires optimizing over all possible measurements, leading to a nonconvex inner problem without a closed form in general.
We therefore define the QMLP given samples $x^n \in \cX^n$  as any minimizer of \begin{align}
    & \inf_{\sigma \in \Sigma} \qrel{\hat \rho_n(x^n)}{\sigma},  \label{eq:mloptrkhsemp} 
\end{align}
whenever the infimum is attained. This formulation has the additional advantage that it leads to convergence guarantees for the MLP in terms of quantum relative entropy, which is a stronger discrepancy measure than measured relative entropy.

   When $\Sigma$ is weakly compact\,\footnote{The compactness assumption  is  used henceforth only to guarantee that the relevant infimum is achieved.}, the infimum in \eqref{eq:mloptrkhsemp} is achieved. We will assume that this is the case throughout the paper.  Moreover, when $\Sigma$ is also convex\,\footnote{Since $\Sigma$ is a weakly compact convex set of density operators, it is easy to see that there exists an element $\sigma \in \Sigma$  such that $\supp(\Sigma)=\sigma$, hence,  $\sigma' \ll \sigma$ for every $\sigma' \in \Sigma$. A standard construction is by taking $\sigma=\sum_{n \in \NN}2^{-n}\sigma_n $ for a countable dense (in trace norm) subset $\{\sigma_n\}_{n \in \NN} $ of $\Sigma$,  which exists as the set of density operators on a separable Hilbert space is separable with respect to trace norm. Hence, one may define $\supp(\Sigma) $ as $ \supp(\sigma)$ for such a $\sigma$.}, 
   $\supp\left(\hat \rho_n(x^n)\right)=\supp(\Sigma)$, and the value of \eqref{eq:mloptrkhsemp} is finite, this minimum is unique. However, the aforementioned support condition is restrictive  and hard to ensure for all $x^n \in \cX^n$.  
Accordingly, we call the  QMLP an arbitrary minimizer of \eqref{eq:mloptrkhsemp}, when the minimizer is not unique. The motivation behind considering  QMLP 
stems from the possibility that with an appropriately chosen embedding $\varphi$ into a Hilbert space $\HH_d$ of dimension $d \ll |\cX|$, $\hat \rho_n(x^n)$ is a good proxy of $\hat p_n$ for the prediction task while simultaneously enabling optimization over a subset of a much smaller Hilbert space $\cL(\HH_d)$, i.e., over the space of  $d \times d$  matrices. This occurs, for instance, when \enquote{neighboring symbols} which should lead to similar prediction outcomes are embedded to closely aligned elements in the Hilbert space, so that the span of $\{\varphi(x): x \in \cX\}$ has a low effective dimension compared to orthogonal embedding.  Thus, as will become evident in Section \ref{Section:Statprop}, a useful embedding should balance low effective dimension with good conditioning of the embedded state $\rho_p$.

From an optimization standpoint, \eqref{eq:mloptrkhsemp} can be solved by optimization algorithms over Hilbert spaces that can exploit its rich structure such as orthogonality and inner products, and lack in $\cP(\cX)$. In fact, there is a rich theory of optimization in Hilbert spaces (e.g., \citet{Houska-Chachuat-2017}), and gradient-based algorithms can exploit the aforementioned structure. 
\section{Main Results}
We next obtain performance guarantees for QMLP. 
From \eqref{eq:mloptrkhsemp}, the optimization problem of interest for determining QMLP  is the analogue of \eqref{eq:MLopt} in $\cL(\HH)$, i.e., 
\begin{align}
    \inf_{\sigma \in \Sigma} \qrel{\rho}{\sigma},  \label{eq:mloptrkhs}
\end{align} 
where $\Sigma$ is a non-empty compact subset of $\cS(\HH)$. Note that since $\qrel{\rho}{\sigma}+\mathsf{H}(\rho)=-\tr{\rho \log \sigma}$, \eqref{eq:mloptrkhs}  is equivalent to minimizing the expected log loss for  prediction  when the true state is $\rho$ and the loss incurred by a model  $\sigma \in \Sigma$  is given according to the observable $-\log \sigma$.  When $\Sigma$ is also convex, \eqref{eq:mloptrkhs} is a convex optimization problem owing to the joint convexity of quantum relative entropy in its arguments (see e.g. \citet{wilde2017quantum}).  In the following, we characterize some structural and statistical properties of QMLP.
\subsection{Structural and Statistical Properties of Quantum Maximum Likelihood Predictor}\label{Section:Statprop}
 Solving \eqref{eq:mloptrkhs} involves optimization of quantum relative entropy over a class of density operators. 
 It is of interest to determine conditions under which this optimization  reduces to its relatively simpler classical analogue given in \eqref{eq:MLopt}.  We show that this indeed happens when $\Sigma$ has sufficiently rich structure. 
To state the conditions required precisely, let $\cE_{\rho}\coloneqq \{\ket{e_i(\rho)}\}_{i=1}^{\infty}$ denote an orthonormal eigenbasis of $\rho$ and $\mathsf{P}_i(\rho)\coloneqq \ket{e_i(\rho)}\bra{e_i(\rho)}$ denote the orthogonal projection corresponding to $\ket{e_i(\rho)}$.  For $\Sigma \subseteq \cS(\HH)$, define the following set  obtained by pinching\,\footnote{This is slightly different from standard pinching, where projectors of similar eigenvalues are combined (see e.g., \citet{Hayashi_2002}).} 
the elements of $\Sigma$ by the rank-one orthogonal projections $\{\mathsf{P}_i(\rho)\}_{i=1}^{\infty}$: 
\begin{align}
&\cJ\left(\cE_{\rho},\Sigma\right) \coloneqq \left\{ \sum_{i=1}^{\infty}\mathsf{P}_i(\rho) \sigma \mathsf{P}_i(\rho): \sigma \in \Sigma \right\}. \label{eq:unitprjinvset}  
\end{align}
It is easy to verify that the operators  in  $\cJ\left(\cE_{\rho},\Sigma\right)$ commute with $\rho$.
Whenever $\tau\in\cJ(\cE_{\rho},\Sigma)$, the vector $\lambda_{\tau}$ in expressions such as $\kl{\lambda_{\rho}}{\lambda_{\tau}}$ is indexed according to the common basis $\cE_{\rho}$. 
 Next,  we call a set $\Sigma \subseteq \cL(\HH)$ \textit{unitarily invariant} if for all $\sigma \in \Sigma $ and unitaries $U$,  $U \sigma U^{\dag} \in \Sigma $. 
We have the following proposition that provides conditions on $\Sigma$ under which   QMLP reduces to an MLP.
\begin{prop}[Relation between MLP and QMLP]\label{prop:QMLPrel}
Let   $\rho\in \cS(\HH)$, $\cE_{\rho}$ be any orthonormal eigenbasis of $\rho$, and 
$\Sigma \subseteq \cS(\HH) $ be a non-empty set.  If $\cJ\left(\cE_{\rho},\Sigma\right) \subseteq \Sigma$, then
\begin{align}
  \inf_{\sigma \in \Sigma}\mathsf{D}(\rho \| \sigma)=\inf_{\sigma \in  \cJ\left(\cE_{\rho},\Sigma\right)} \kl{\lambda_{\rho}}{\lambda_{\sigma}}.\label{eq:mlpexctexp}
\end{align} 
If $\Sigma$ is also unitarily invariant, then the above terms also equal $\inf_{\sigma \in \Sigma} \kl{\lambda_{\rho}}{\lambda_{\sigma}}$.
\end{prop}
The proof of Proposition \ref{prop:QMLPrel} (see Section \ref{Sec:prop:QMLPrel-proof}) is based on elementary arguments involving data processing inequality for quantum relative entropy, quantum measurements and the unitary invariance of $\Sigma$.
\begin{remark}[Relevance of general QMLP]
    Although Proposition \ref{prop:QMLPrel} shows that QMLP reduces to an MLP for a sufficiently rich $\Sigma$, the conditions therein on $\Sigma$ are hard to be satisfied in practice (when solving \eqref{eq:mloptrkhs} on a classical or quantum computer). For instance, it is well known based on a volumetric  argument  that the number of basic quantum gates required for implementing a covering of the unitary group scales super-exponentially in the number of qubits \citep{Nielsen_Chuang_2010}. While additional symmetry assumptions on the class such as permutation invariance can be exploited to reduce the super-exponential dependence to polynomial (see e.g.  \citep[Section 4]{SGW-2026}), this still only achieves an $\epsilon$-covering of the unitary group and not unitary invariance.  Hence, the general QMLP formulation is still relevant. 
\end{remark}
Having seen some structural aspects of QMLP, we next study its statistical behaviour. 
The  following technical proposition  will play a key role towards this purpose. 
\begin{prop}[QMLP distance bound] \label{Prop:devmlpred}
     Let  $\rho,\tilde \rho \in \cS(\HH)$ and  $\Sigma \subseteq \cS(\HH)$ be compact. Let 
\begin{align}
     \sigma^{\star} \coloneqq \argmin_{\sigma \in \Sigma} \mathsf{D}(\rho \| \sigma) \quad \text{and}\quad 
     \tilde{\sigma}^{\star}  \coloneqq \argmin_{\sigma \in \Sigma} \mathsf{D}(\tilde{\rho} \| \sigma). \label{eq:argminempdist}
    \end{align}
If $\Sigma$ satisfies $ \mathsf{D}(\rho \| \sigma^{\star}) \leq \epsilon$ for some $\epsilon \geq 0$, then 
\begin{subequations}\label{eq:trdistdenop}
    \begin{align}
    \norm{ \tilde{\sigma}^{\star} -\sigma^{\star}}_1^2 &\leq 4\big(\qrel{\tilde \rho}{\sigma^{\star}}-\qrel{\rho}{\sigma^{\star}}\mspace{-3 mu}\big) +4 \norm{\tilde \rho-\rho}_1^2+12\epsilon, \label{eq:trdistdenop1} \\
    \norm{\tilde{\sigma}^{\star}-\rho}_1 &\leq \norm{ \tilde{\sigma}^{\star}-\sigma^{\star}}_1 + \sqrt{2\epsilon}. \label{eq:trdistdenop2} 
\end{align} 
\end{subequations}
Additionally, if $\Sigma$ is unitarily invariant and  $\cJ\left(\cE_{\rho},\Sigma\right) \subseteq \Sigma $ (see  \eqref{eq:unitprjinvset}) for some orthonormal eigenbasis $\cE_{\rho}$ of $\rho$, then \eqref{eq:trdistdenop1} holds with $\qrel{\tilde \rho}{\sigma^{\star}}-\qrel{\rho}{\sigma^{\star}}$ replaced by $\kl{\lambda_{\tilde \rho}}{\lambda_{\sigma^{\star}}}-\kl{\lambda_{\rho}}{\lambda_{\sigma^{\star}}}$.
\end{prop}
Proposition \ref{Prop:devmlpred}    is proved (see Section \ref{Sec:Prop:devmlpred-proof}) based on elementary arguments involving quantum Pinsker's inequality \citep[Equation 3.53]{Hayashi-book-2016}.  We will use it to prove Theorem \ref{Thm:convergence-MLpred} which provides statistical non-asymptotic performance guarantees for QMLP.

 To this end, we specialize to embeddings into $\HH_d$ for some $d \in \NN$. Embeddings into finite dimensional Hilbert spaces are practically important, since  LLMs and QLLMs are based on such embeddings.  For $\rho,\sigma >0$, let the Thompson metric \citep{thompson_1963} on the space of positive density operators be  
 $T(\rho,\sigma)\coloneqq \log \!\big(\big\|\sigma^{-1/2}\rho \sigma^{-1/2} \big\|\vee  \big\|\rho^{-1/2} \sigma \rho^{-1/2}\big\|\big)$. 
 Note that $T(\rho,\sigma)=0$ if and only if $\rho=\sigma$. 
  
 For technical reasons, we will consider a slightly perturbed version of $\hat \rho_n(x^n)$ obtained using the maximally mixed state $\pi_d$ ($\coloneqq I/d$ where $I$ denotes the identity operator) in $\HH_d$ such that the perturbation vanishes asymptotically with $n$. Accordingly, define 
  \begin{align}
      \rho_n(x^n) \coloneqq \left(1-\frac{1}{n}\right) \hat \rho_n(x^n) +\frac{1}{n} \pi_d, \label{eq:empMLdist}
  \end{align}
  where $\hat \rho_n(x^n)$ is as defined in \eqref{eq:denopemb}, and set
\begin{align}\hat{\sigma}_n^{\star}(x^n)   \coloneqq \left(1-\frac{1}{n}\right)  \sigma_n^{\star}(x^n) +\frac{1}{n} \pi_d \quad \mbox{with } \quad 
 \sigma_n^{\star}(x^n)   \coloneqq \argmin_{\sigma \in \Sigma} \mathsf{D}(\rho_n(x^n)\| \sigma). \label{eq:empMLpred}
  \end{align}
Note that the definition in \eqref{eq:empMLdist} involves a perturbation of $\hat \rho_n(x^n)$ with $\pi_d$, which requires knowledge of $\HH_d$. However, this can be determined given the embedding $\varphi$ and knowledge of the support of $P$, as the span of $\{\varphi(x):x \in \cX,\,p(x)>0\}$.

 The next theorem establishes consistency and convergence properties of  QMLP. 
\begin{theorem}[Consistency and concentration  of QMLP]\label{Thm:convergence-MLpred}
Let  $\rho_p>0$ and  $\rho_n(x^n)>0$ be density operators on $\HH_d$  as defined in \eqref{eq:covembed}   and \eqref{eq:empMLdist}, respectively. 
Suppose  $ \Sigma$ is a compact  set with $\supp(\Sigma)=\HH_d$ such that 
$\sigma_p^{\star} \coloneqq \argmin_{\sigma \in \Sigma} \mathsf{D}(\rho_p \| \sigma)$ satisfies  
    $\qrel{\rho_p}{\sigma_p^{\star}} \leq \epsilon$. 
Set \begin{align}
  &  b_n\coloneqq \log \norm{\sigma_p^{\star-1}} \vee \log (dn) \vee T(\rho_p, \sigma_p^{\star}), \quad  \bar{b}_n \coloneqq \log (dn) \vee \log \norm{\rho_p^{-1}},\notag \\
  \text{ and } &  \delta_n\coloneqq 0.5 \norm{\rho_p^{-1}}-n^{-1}.  \label{eq:bndfuncclls} 
\end{align}
Then, with $ \sigma_n^{\star}(x^n)$ and $\hat{\sigma}_n^{\star}(x^n)$ as defined in \eqref{eq:empMLpred}, the following hold for  $X^n \sim P^{\otimes n}$:
\begin{enumerate}[(i)]
    \item  \textbf{Performance guarantees under (squared) trace norm:}
     \begin{align} 
       & \EE\big[ \norm{\sigma_n^{\star}(X^n)-\rho_p}_1^2\big] \leq \begin{cases}
         16d(b_n+4)\big(\sqrt{2\log (2d)}+\sqrt{\pi/2}\big) n^{-\frac 12}  \\
         \qquad \qquad  + 16b_n n^{-1}+64 n^{-2}+28\epsilon, & \text{ if } \epsilon>0,  \\
         (32 d \norm{\rho_p^{-1}}+ 8 d^2) \left(\frac{8}{n^2} + \frac{28d}{n}\right), & \text{ if }\epsilon=0. 
     \end{cases} \label{eq:conv-rate}
    \end{align}   
Moreover, for all $ t \geq 0$, we have
\begin{subequations}\label{eq:conc-ineq-comb}
     \begin{align}
     & \PP\left(\norm{\sigma_n^{\star}(X^n)-\rho_p}_1^2 \geq \frac{16 b_n}{n} +\frac{64}{n^2}+28\epsilon+8d (b_n+4)t\right) \mspace{-3 mu} \leq 
          4d e^{-\left(\frac{n t^2}{4} \wedge \frac{3nt}{4}\right)}, \mbox{ if } \epsilon>0, \label{eq:devineqmlpred} \\
           &  \PP\left(\norm{\sigma_n^{\star}(X^n)-\rho_p}_1^2 \geq   (32 d \norm{\rho_p^{-1}}+ 8 d^2)\left(\frac{8}{n^2}+2t\right)\right) \leq 2d e^{-\left(\frac{n t}{4} \wedge \frac{3n\sqrt{t}}{4}\right)}, \text{ if } \epsilon=0.\label{eq:conc-ineq}
           \end{align}
    \end{subequations}
\item \textbf{Performance guarantees under quantum relative entropy:}
 If $\Sigma$ is such that 
\begin{align}
 \EE\left[\qrel{\rho_n(X^n)}{\sigma_n^{\star}(X^n)}\right] \wedge \EE\left[\qrel{\rho_n(X^n)}{\hat{\sigma}_n^{\star}(X^n)}\right] \leq \epsilon, \label{eq:klassump}    
\end{align}
then     
  \begin{align}
       & \EE\big[ \qrel{\rho_p}{\hat{\sigma}_n^{\star}(X^n)}\big] \mspace{-4 mu} \notag \\
        &\leq \begin{cases}
           \mspace{-3 mu}(2\bar{b}_n+\log d) n^{-1}\mspace{-2 mu}+ d \bar{b}_n\mspace{-4 mu}\left(\mspace{-4 mu}\sqrt{2\log (2d)}+\sqrt{\pi/2}\right) n^{-\frac 12}+\epsilon, &\text{if } \epsilon>0, \label{eq:experrorqrel} \\
          \mspace{-3 mu}\frac{2\norm{\rho_p^{-1}}+\log d}{n}+2d\log(dn)e^{-\left(\frac{n\delta_n^2}{4}\wedge\frac{3n\delta_n}{4}\right)} &\text{if } \epsilon=0\text{ and }n>2\norm{\rho_p^{-1}}
       \end{cases}
       \end{align}
   If $\Sigma$ also satisfies     
    \begin{align}
        \qrel{\rho_n(X^n)}{\sigma_n^{\star}(X^n)} \wedge \qrel{\rho_n(X^n)}{\hat{\sigma}_n^{\star}(X^n)} \leq \epsilon, \quad \text{almost surely,} \label{eq:asqrelbnd}
    \end{align}
 then  for all $t \geq 0$,    
 \begin{subequations}
 \begin{align}
& \PP\left( \qrel{\rho_p}{\hat{\sigma}_n^{\star}(X^n)} \geq \frac{2\bar{b}_n}{n}+\frac{\log d}{n}+\epsilon+d \bar{b}_n t  \right)  \leq 2d e^{-\left(\frac{n t^2}{4 } \wedge \frac{3nt}{4 }\right)},\quad \text{if } \epsilon>0, \label{eq:devineqqrel} \\
&\PP\left( \qrel{\rho_p}{\hat \sigma_n^{\star}(X^n)} \geq 2d\norm{\rho_p^{-1}}\left(t+\frac1n\right)^2+\frac{\log d}{n}  \right)  \leq 2d e^{-\left(\frac{n t^2}{4} \wedge \frac{3nt}{4}\right)}, \notag \\
&\hspace{8.7cm}\text{if } \epsilon=0,\ n>2\norm{\rho_p^{-1}},\ 0\leq t\leq\delta_n.\label{eq:devineqqrelmatched}
\end{align}\label{eq:devineqmatchmism}
\end{subequations}
\end{enumerate}
\end{theorem}
The proof of Theorem \ref{Thm:convergence-MLpred}  (see  Section \ref{Sec:Thm:convergence-MLpredKL-proof}) combines the variational form of quantum relative entropy \eqref{eq:varexpqrelent} with Proposition \ref{Prop:devmlpred} to control $\norm{\sigma_n^{\star}(X^n)-\rho_p}_1$ and $\qrel{\rho_p}{\hat{\sigma}_n^{\star}(X^n)}$, and then applies matrix Bernstein or Hoeffding bounds.
 Observe that \eqref{eq:conv-rate} implies a parametric rate of convergence with respect to $n$ for the QMLP to the target distribution (up to the approximation factor $\epsilon$) when $\epsilon>0$ and $O_d(1/n)$ rate of convergence when $\epsilon=0$. Hence, there is no  curse of dimensionality in the convergence rate. Moreover,   given the additional condition  in \eqref{eq:klassump} holds, the corresponding rates of convergence under quantum relative entropy are $\tilde O(d/\sqrt{n})$ when $\epsilon>0$  and $O\big(\big(\norm{\rho_p^{-1}}+\log d\big)/n\big)$ when $\epsilon=0$, respectively.   It can also be seen that \eqref{eq:experrorqrel} and \eqref{eq:devineqmatchmism}, combined with Pinsker's inequality, yield the analogues of \eqref{eq:conv-rate} and \eqref{eq:conc-ineq-comb}, respectively,  with $\norm{\sigma_n^{\star}(X^n)-\rho_p}_1$ replaced by $\norm{\hat{\sigma}_n^{\star}(X^n)-\rho_p}_1$. Further, by data processing inequality for trace norm and quantum relative entropy, the claims in Theorem \ref{Thm:convergence-MLpred} also hold for the output probability distributions obtained after processing the QMLP and  the embedded target distribution $\rho_p$.   

A few remarks about the requirements for Theorem \ref{Thm:convergence-MLpred} to hold are in order.
The assumption $\rho_p>0$ is not very restrictive since $\hat \rho_n(X^n) \ll   \rho_p$ almost surely (the complementary event requires $X_i \notin \supp(P) $ for some $1 \leq i \leq n$, which has probability zero), and the support of $\Sigma$  can be restricted to support of $\rho_p$ which can be determined via knowledge of $\supp(P)$ given an  embedding.  
Moreover, in a finite-dimensional Hilbert space $\HH_d$, a set $\Sigma$ satisfying the conditions in Theorem \ref{Thm:convergence-MLpred} exists (trivially). For instance,  $\Sigma=\cS(\HH_d)$ is a  compact  set satisfying \eqref{eq:klassump} and \eqref{eq:asqrelbnd}. More generally, any $\epsilon$-covering of $\cS(\HH_d)$ in  quantum relative entropy  as a measure of separation suffices. Here, an $\epsilon$-covering is a set $\cA \subseteq \cS(\HH_d)$ such that for any $\rho \in \cS(\HH_d)$, there exists a $\sigma \in \cA$ with $\qrel{\rho}{\sigma} \leq \epsilon$.  Let  $N\big(\epsilon, \cS(\HH_d), \qrel{\cdot}{\cdot}\mspace{-2 mu}\big)$ denote  the minimum cardinality of such a set $\cA$, which is referred to as the covering number of $\cS(\HH_d)$ in quantum relative entropy. Since we were unable to find concrete bounds in the literature for this quantity, the next proposition provides an upper bound  for the same.
\begin{prop}[Covering number bounds in quantum relative entropy]\label{Prop:covnumdensop}
There exists $0 <c < 65$ such that
\begin{align}
    N\big(\epsilon, \cS(\HH_d), \qrel{\cdot}{\cdot}\big) \leq \left(\frac{c d}{\epsilon}\right)^{d^2+d-1}, ~~~\forall~~0<\epsilon \leq 1.
\end{align}
\end{prop}
The proof of Proposition \ref{Prop:covnumdensop} (see Section \ref{Sec:Prop:covnumdensop-proof})  relies on upper bounding quantum relative entropy in terms of a version of quantum chi-square divergence \citep{Temme-2010}, which is a relatively easier quantity to handle.  Then, a covering of $\cS(\HH_d)$ in terms of quantum chi-square divergence can be obtained in terms of a covering of the unitary group and that of the probability simplex in  KL divergence, the latter of which has been studied in detail  in \citet{tang2022divergence}. In particular, we use a covering of the simplex using uniformly spaced centers and obtain  a covering of the unitary group using Euclidean covering like arguments.

 Theorem \ref{Thm:convergence-MLpred} provides insights on certain properties that a good embedding should satisfy. Specifically, a good embedding $\varphi$ should be such that the minimal eigenvalue of $\rho_p$ is as high  as  possible subject to satisfactory prediction performance. Another objective is to minimize effective dimension by embedding neighboring symbols (with similar prediction outcomes) to  vectors that are closely aligned. 
 In contrast, classical information-theoretic analyses for prediction obtains performance guarantees by considering only the source probability distribution without taking account of similarity or context, which corresponds to the scenario of orthogonal embedding. 

 \noindent
\textbf{Classical Prediction from QMLP}: Let $\cX$ and $\mu$ be such that $\mu(\cX) <\infty$. Then, 
given a measurable embedding $\varphi:\cX \rightarrow \HH_d$ and  QMLP $\hat{\sigma}_n^{\star}$, a natural classical predictor can be obtained by measuring QMLP using the POVM density $\cM_{\varphi}(\cX)\coloneqq \{M_{\varphi}(x),~x \in \cX\}$, where 
\begin{align}
   M_{\varphi}(x)\coloneqq \widebar{M}_{\varphi}^{-\frac{1}{2}} \ket{\varphi(x)}\bra{\varphi(x)}\widebar{M}_{\varphi}^{-\frac{1}{2}}, \quad \mbox{with} \quad \widebar{M}_{\varphi} \coloneqq \int_{\cX} \ket{\varphi(x)}\bra{\varphi(x)}d\mu(x). \notag 
\end{align}
Note that since we may assume without loss of generality that  $\widebar{M}_{\varphi}$ is full rank by restricting the underlying Hilbert space to its support,  $M_{\varphi}(x)$ is a bounded positive operator for each $x \in \cX$  and $\int_{x \in \cX}M_{\varphi}(x) d\mu(x)=I$, and hence a valid POVM density. 
Then, for the pdfs $  \hat q_n(x) \coloneqq \tr{M_{\varphi}(x)\,\hat{\sigma}_n^{\star}}$ and $  \hat p(x) \coloneqq \tr{M_{\varphi}(x)\,\rho_p}$  obtained via measurements using $\cM_{\varphi}(\cX)$, we have via data processing inequality for trace norm that
\begin{align}
  \big\|\hat P-\hat Q_n\big\|_1 =  \norm{\hat p-\hat q_n}_{1,\mu} \leq \norm{\hat{\sigma}_n^{\star}-\rho_p}_1, \notag
\end{align}
where $\hat P$ and $\hat Q_n$ denote the probability measures corresponding to  densities $\hat p$ and $\hat q_n$, respectively.
Similarly, data processing inequality for quantum relative entropy yields $   \mathsf{D}_{\mathsf{KL}}(\hat P\|\hat Q_n) \leq \qrel{\rho_p}{\hat{\sigma}_n^{\star}}$. 
Hence, proximity of QMLP to the image of the true distribution in the embedded domain as quantified by Theorem \ref{Thm:convergence-MLpred} also translates to the closeness of  classical probability distributions obtained via their respective measurements.
 \subsection{Quantum Information Projection}
Next, we revisit the problem of \textit{information projection} (or $I-$projection) \citep{csiszar2003information, Csiszar-Shields-2004} in the quantum setting. This is the counterpart of  QMLP (also called \textit{quantum reverse  $I-$projection}) in quantum information geometry \citep{Amari-Nagaoka-2000,Hayashi-book-2016}, where  quantum relative entropy is optimized over a class of quantum states  with respect to its \textit{first} argument. 
Formally, the $I-$projection  of $\sigma \in \cS(\HH)$ into $\bar{\cS} \subseteq \cS(\HH) $ is 
\begin{align}
\rho^{\star}_{\sigma,\bar{\cS}}\coloneqq \argmin_{\rho \in \bar{\cS}} \qrel{\rho}{\sigma}, \label{eq:mininfproj} 
\end{align}
if it exists (any minimizer if multiple ones exist). As before, we will consider that $\bar{\cS} $ is a non-empty compact convex set. In this case,    $\rho^{\star}_{\sigma,\bar{\cS}}$ exists owing to lower semicontinuity of $\qrel{\rho}{\sigma}$ in  $\rho$. Moroever, it is  unique if $\bar{\cS} \cap \{\rho \in \cS(\HH):\qrel{\rho}{\sigma}<\infty\}$  is non-empty due to strict convexity of  $\qrel{\rho}{\sigma}$ restricted to this set. 
An important result in quantum information geometry involving the $I$-projection is the 
 \textit{quantum Pythagorean theorem}  (see e.g., \citet{Amari-Nagaoka-2000,petz2007quantum,Hayashi-book-2016,Hayashi-Ito-2025}). In the following, we provide some novel generalizations of this theorem in finite and infinite dimensions. 
To state our results, we need to introduce a few notions. Call a  set $\cS \subseteq \cS(\HH)$ of density operators \textit{linearly closed} if $\alpha \rho +(1-\alpha) \tilde \rho  \in \cS$  for every $\rho, \tilde \rho \in \cS$ and $\alpha \in \RR$ such that $\alpha \rho +(1-\alpha) \tilde \rho \in \cS(\HH)$. 
Denote the set of bounded linear operators by $\cB(\HH)$, and let $\cC=\{L_i\}_{i=1}^k \subseteq \cB(\HH)$ be a finite set.  For a given $\rho_0 \in \cS(\HH)$ and $\cC$, the set of all density operators $\rho \in \cS(\HH)$ such that 
$\tr{\rho L_i}=\tr{\rho_0 L_i}$, for  $1 \leq i \leq k$, is called a \textit{mixture} or \textit{linear} family generated by $\cC$ (and $\rho_0$), and denoted by $\cM(\rho_0,\cC)$. Defining
\begin{align}
\cC_{\mathsf{H}}\coloneqq\left\{\frac{L_i+L_i^{\dagger}}{2},\frac{L_i-L_i^{\dagger}}{2\mathrm i}:1\leq i\leq k\right\}, \label{eq:Hermitian-closure}
\end{align}
the complex constraints defining $\cM(\rho_0,\cC)$ are equivalent to the real constraints generated by $\cC_{\mathsf{H}}$, so $\cM(\rho_0,\cC)=\cM(\rho_0,\cC_{\mathsf{H}})$.
For a sub-density operator $\sigma_0$, i.e.,  $\sigma_0 \geq 0$ and $\tr{\sigma_0}\in(0,1]$,  and   $\cC$, 
the set of all density operators of the form 
\begin{align}
    \sigma= \frac{e^{\log \sigma_0+\sum_{i=1}^k \beta_i \mathsf{P}_{\sigma_0}L_i\mathsf{P}_{\sigma_0} }}{c(\sigma_0,\bm{\beta},\cC)}, ~ \text{with} ~ \bm{\beta}=(\beta_1,\ldots,\beta_k) \in \CC^k, ~ c(\sigma_0,\bm{\beta},\cC) \coloneqq \tr{e^{\log \sigma_0 +\sum_{i=1}^k \beta_i \mathsf{P}_{\sigma_0}L_i\mathsf{P}_{\sigma_0}}}, \notag 
\end{align}
is called  \textit{exponential} family generated by $\cC$ (and $\sigma_0$), and denoted by $\mathrm{Exp}(\sigma_0,\cC)$. Here,  $c(\sigma_0,\bm{\beta},\cC)$
is the normalization factor to make $\tr{\sigma}=1$, $\mathsf{P}_{\sigma_0}$ denotes the orthogonal projection onto $\supp(\sigma_0)$,  and $e^{\log \sigma_0+\sum_{i=1}^k \beta_i \mathsf{P}_{\sigma_0}L_i \mathsf{P}_{\sigma_0} }$ is considered as an operator on $\supp(\sigma_0)$. Note that  since $\cC$ can contain non self-adjoint operators, and so not all  $\bm{\beta}$ in the definition of exponential family corresponds to a density operator. Specifically, $\bm{\beta}$ and $\cC$ have to be such that $\sum_{i=1}^k \beta_i \mathsf{P}_{\sigma_0} L_i \mathsf{P}_{\sigma_0}$ is self-adjoint.   
Also, observe  that a linearly closed set is convex by definition and  mixture families are linearly closed. On the other hand, the exponential family is not closed nor linearly closed.

\begin{figure}[t]
\centering
\includegraphics[trim=0cm 0cm 0cm 0cm, clip, width= 0.6\textwidth]{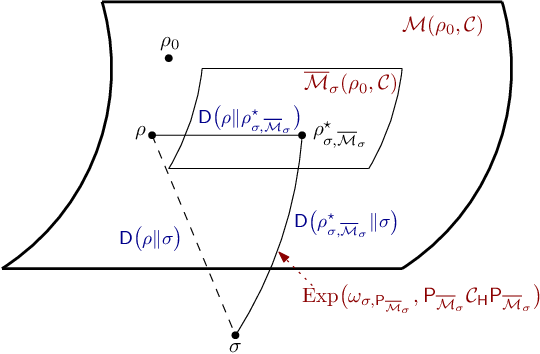}
\caption{Illustration of quantum Pythagorean theorem} \label{fig:I_proj} 
\end{figure}

We can now state a  version of our quantum Pythagorean theorem in the setting of a finite dimensional Hilbert space $\HH_d$ (see  Figure \ref{fig:I_proj} for an illustration).
\begin{theorem}[Quantum Pythagorean theorem in finite dimensions]\label{Thm:quantPyth}
Consider a finite dimensional Hilbert space $\HH_d$. Let $\sigma \in \cS(\HH_d)$ and $\bar{\cS} \subseteq \cS(\HH_d)$ be a non-empty compact convex set. Then, the following hold:
\begin{enumerate}[(i)]
    \item If $\supp(\bar{\cS}) \subseteq \supp(\sigma)$, the unique $I$-projection $\rho^{\star}_{\sigma,\bar{\cS}}$ satisfies $\supp(\rho^{\star}_{\sigma,\bar{\cS}})=\supp(\bar{\cS})$ and
\begin{align}
    \qrel{\rho}{\sigma} \geq \mathsf{D}\big(\rho\|\rho^{\star}_{\sigma,\bar{\cS}}\big)+\mathsf{D}\big(\rho^{\star}_{\sigma,\bar{\cS}}\|\sigma\big), \quad \forall~\rho \in \bar{\cS}, \label{eq:pythgthm}
\end{align}
with equality holding when $\bar{\cS}$ is linearly closed.

\item For a finite set $\cC=\{L_i\}_{i=1}^k \subseteq \cB(\HH_d)$, let $\cM(\rho_0,\cC)$ be the mixture family generated by $\cC$. Let $\widebar{\cM}_{\sigma}\coloneqq\{\rho\in\cM(\rho_0,\cC):\rho\ll\sigma\}$ be non-empty, and with  $\mathsf{P}\coloneqq \mathsf{P}_{\widebar \cM_{\sigma}}$,  define the density operator
\begin{align}
\omega_{\sigma,\mathsf{P}}\coloneqq
\frac{\exp(\mathsf{P}\log\sigma \mathsf{P})}{\tr{\exp(\mathsf{P}\log\sigma \mathsf{P})}}. \label{eq:expfam-reference}
\end{align}
Then, defining  $\mathsf{P}\cC_{\mathsf{H}}\mathsf{P}\coloneqq\{\mathsf{P}K\mathsf{P}:K\in\cC_{\mathsf{H}}\}$, where $\cC_{\mathsf{H}}$ is as defined in \eqref{eq:Hermitian-closure},
\begin{align}
& \rho^{\star}_{\sigma,\widebar{\cM}_{\sigma}}
\in \mathrm{Exp}\big(\omega_{\sigma,\mathsf{P}},\mathsf{P}\cC_{\mathsf{H}} \mathsf{P}\big)\cap\widebar{\cM}_{\sigma}, \label{eq:expfam} \\
 \text{and} \qquad    &\qrel{\rho}{\sigma} = \mathsf{D}\big(\rho\|\rho^{\star}_{\sigma,\widebar{\cM}_{\sigma}}\big)+\mathsf{D}\big(\rho^{\star}_{\sigma,\widebar{\cM}_{\sigma}}\|\sigma\big), \quad \forall~\rho \in \cM(\rho_0,\cC). \label{eq:pythgthmmix}
\end{align}

\item Let $\rho,\sigma \in \cS(\HH_d)$ be such that $\rho\ll\sigma>0$, and let $\cE_{\rho}$ be an orthonormal eigenbasis of $\rho$. Set $\bar{\cS}(\cE_{\rho})$ to be the set of density operators diagonal in $\cE_{\rho}$. Then \eqref{eq:pythgthmmix} holds with
\begin{align}
\rho^{\star}_{\sigma,\bar{\cS}(\cE_{\rho})}
=\frac{\exp\!\left(\sum_{i=1}^d \mathsf{P}_i(\rho)\log\sigma\,\mathsf{P}_i(\rho)\right)}
{\tr{\exp\!\left(\sum_{i=1}^d \mathsf{P}_i(\rho)\log\sigma\,\mathsf{P}_i(\rho)\right)}}. \label{eq:diag-Iproj-corrected}
\end{align}
\end{enumerate}
\end{theorem}
\begin{remark}[Support conditions in Theorem \ref{Thm:quantPyth}]\label{rem:suppcond}
If $\supp(\bar{\cS})\not\subseteq\supp(\sigma)$, set $\bar{\cS}_{\sigma}\coloneqq\{\rho\in\bar{\cS}:\rho\ll\sigma\}$. If $\bar{\cS}_{\sigma}$ is non-empty, Part $(i)$ applies to $\bar{\cS}_{\sigma}$, and the minimizer over $\bar{\cS}$ lies in $\bar{\cS}_{\sigma}$. If $\bar{\cS}_{\sigma}$ is empty, then both sides of \eqref{eq:pythgthm} are infinite and any $\rho \in \bar{\cS}$ could be taken as $\rho^{\star}_{\sigma,\bar{\cS}}$.  For $\rho\in\bar{\cS}\setminus\bar{\cS}_{\sigma}$, $\mathsf D(\rho\|\sigma)=\infty$.
\end{remark}
 Theorem \ref{Thm:quantPyth} (see Section \ref{Sec:Thm:quantPyth-proof}) extends   \citet[Theorem 3.2]{Csiszar-Shields-2004}  to the non-commutative setting, carefully accounting for support conditions.  The proof of the main claim in Part $(i)$ relies on  Donald's identity for quantum relative entropy \citep[Section 2]{Donald1987FreeEA}.    Part $(ii)$, which is a special case of  Part $(i)$, considers a more general  exponential and mixture family (generated by bounded linear operators) than we are aware of in the quantum literature. This generalization is used in the proof of Part $(iii)$, which involves exponential and mixture families generated by the class $\cC=\{\ket{e_i(\rho)}\bra{e_j(\rho)} \}_{1 \leq i \neq j \leq d}$, where $ \{e_i(\rho)\}_{i=1}^{d}$ denotes an orthonormal eigenbasis of $\rho$. It is also interesting that the $I-$projection of $\sigma>0 $ to $\bar{\cS}(\cE_{\rho})$  is simply characterized as the exponential of the pinched operator $\sum_{i=1}^d \mathsf{P}_i(\rho)\,\log \sigma\, \mathsf{P}_i(\rho)$,  normalized to be a density operator.

An analogue of the Pythagorean  inequality in \eqref{eq:pythgthm} can be obtained in the  setting of an infinite dimensional Hilbert space.
 \begin{theorem}[Quantum Pythagorean inequality -- infinite dimensional]\label{Thm:Qpyth-infdim}
Let $\sigma\in\cS(\HH)$ and let $\bar{\cS}\subseteq\cS(\HH)$ be a non-empty compact convex set such that
\begin{align}
\inf_{\rho\in\bar{\cS}}\qrel{\rho}{\sigma}<\infty. \label{eq:infdim-finite-inf}
\end{align}
Let $\hat{\cS}\mspace{-3 mu}\coloneqq \mspace{-3 mu}\{\rho\in\bar{\cS}:\qrel{\rho}{\sigma}\mspace{-2 mu}<\mspace{-2 mu}\infty\mspace{-1 mu}\}$. Then the unique minimizer $\rho^{\star}_{\sigma,\bar{\cS}}$ belongs to $\hat{\cS}$ and, for every $\rho\in\bar{\cS}$,
\begin{align}
\qrel{\rho}{\sigma}\geq
\mathsf D\big(\rho\|\rho^{\star}_{\sigma,\bar{\cS}}\big)
+\mathsf D\big(\rho^{\star}_{\sigma,\bar{\cS}}\|\sigma\big).  \label{eq:pythgthm-infdim}
\end{align}
Moreover, $\supp(\rho^{\star}_{\sigma,\bar{\cS}})=\supp(\hat{\cS})$ and $\mathsf D(\rho\|\rho^{\star}_{\sigma,\bar{\cS}})<\infty$ for every $\rho\in\hat{\cS}$. 
\end{theorem}
The proof of Theorem \ref{Thm:Qpyth-infdim} (see Section \ref{Sec:Thm:quantPyth-proof}) again utilizes Donald's identity along with lower semicontinuity of quantum relative entropy. We remark that a general information projection  inequality in the framework of von Neumann algebras have been obtained by \citet{Jencova-2005} for the family of $\alpha$-divergences with $\alpha \in (-1,1)$. However, quantum relative entropy corresponds to the limiting singular case $\alpha=1$ in this parametrization, which is not covered directly by those results.  

\begin{remark}[Assumptions in Theorem \ref{Thm:Qpyth-infdim}]\label{Rem:AssumpinfdimPyththm}
Compactness is used only to guarantee existence of a minimizer. More generally, the same conclusion holds for any convex set $\bar{\cS}$ for which the  infimum in \eqref{eq:infdim-finite-inf} is finite and  attained at a state $\rho^{\star}_{\sigma,\bar{\cS}}$.
\end{remark}
\subsection{Computing  Quantum Maximum Likelihood Predictor}
We focused in this paper on statistical and geometrical aspects of QMLP. An important practical aspect,  the computation of the QMLP, is an optimization problem that has many applications in quantum  information theory, quantum statistics, quantum machine learning, and quantum resource theories, and has been discussed in various papers.  For instance, the QMLP optimization problem appears in  quantifying the relative entropy of entanglement, Gibbs/Hamiltonian learning, Boltzmann machines, generative modeling and resource conversion, to name a few. Therefore, here, we briefly discuss some of the available methods  for completeness. 

Existing methods for computing QMLP can be largely grouped into  gradient-projection based methods, semidefinite programming,  interior-point methods and variational quantum algorithms. The exact choice of the algorithm depends on the specific application at hand, and what structural constraints on  $\Sigma$ it enforces. Concretely, we may recall that quantum relative entropy is a convex function of its arguments. Hence, if $\Sigma$ is a
convex subset of $\cS(\HH_d)$, computing QMLP is a convex optimization problem. An important instance of this problem pertains to the case when $\Sigma$ is defined by affine and conic constraints. For solving such problems,  several implementations  of solvers based on interior-point methods are already available in the literature \citep{he2025qicsquantuminformationconic,Karimi2023DomainDrivenS}. Such solvers rely on self-concordance of the natural barrier function of quantum relative entropy cone \citep{Fawzi-Saunderson-2023} and are efficient typically  when $d$ is relatively small or moderate, although they can be made to scale well when the  Hessian of the barrier function and the corresponding  Newton equations can be solved efficiently \citep{HeSaundersonFawzi2026,Karimi-Tuncel-2025}.

Methods based on semidefinite programming  have also been considered, e.g.,  for computing the relative entropy of entanglement  via linear approximation of the set of separable states \citep{Zinchenko-2010}, for computing the quantum relative entropy cone via semidefinite approximations of matrix logarithm \citep{Fawzi2018MatrixLog}, and for optimizing relative entropy via semidefinite approximations with computable optimality bounds \citep{KossmannSchwonnek2026}.  Another alternative method is via mirror entropic descent \citep{you2022minimizing}, which has been used recently for computing various quantum information-theoretic quantities. For quantum exponential-family models, quantum iterative-scaling and related first-order methods have been proposed for minimizing reverse quantum relative entropy \citep{GaoJiWei2024}. Closely related  quantum relative-entropy minimization problems also arise in quantum machine learning; in particular, training quantum Boltzmann machines with visible and hidden units can be formulated through the minimization of quantum relative entropy between a target state and a parametrized model state \citep{Wilde2025fundamentals}. First-order Bregman-proximal and primal--dual methods offer yet another complementary method for larger-scale problems where the cost of interior-point iterations may become prohibitive \citep{HeSaundersonFawzi2024}.

\section{Proofs}
\subsection{Proof of Proposition \ref{prop:QMLPrel}}\label{Sec:prop:QMLPrel-proof}
Note that performing a quantum measurement in the eigenbasis $\cE(\rho)$ of $\rho$ and $\sigma$ yields the pmfs $\lambda_{\rho}$ and $\lambda_{\sigma'}$ for $\sigma'=\sum_{i=1}^{\infty}\mathsf{P}_i(\rho) \sigma \mathsf{P}_i(\rho)$. Then, data processing inequality for quantum relative entropy yields
\begin{align}
\mathsf{D}(\rho \| \sigma) \geq  \kl{\lambda_{\rho}}{\lambda_{\sigma'}}. \notag 
\end{align}
Hence
\begin{align}
  \inf_{\sigma \in \Sigma}\mathsf{D}(\rho \| \sigma)\geq  \inf_{\sigma \in  \cJ\left(\cE_{\rho},\Sigma\right)} \kl{\lambda_{\rho}}{\lambda_{\sigma}}.\label{eq:lwrbndmlpval}
\end{align}
Since the opposite inequality holds when $\cJ\left(\cE_{\rho},\Sigma\right) \subseteq \Sigma$ (by using that infimum can only increase when taken over a subset), \eqref{eq:mlpexctexp} follows. 

Let $\Sigma(\cE_{\rho})$ denote the set of density operators in $\Sigma$ for which $\cE_{\rho}$ is an eigenbasis. 
If  $\cJ\left(\cE_{\rho},\Sigma\right) \subseteq \Sigma$ and $\Sigma$ is unitarily invariant, then the last claim in the proposition holds as
\begin{align}
\inf_{\sigma \in \Sigma}\mathsf{D}(\rho \| \sigma) \leq  \inf_{\sigma \in \Sigma(\cE_{\rho})} \mathsf{D}(\rho \| \sigma) =\inf_{\sigma \in \Sigma(\cE_{\rho})} \kl{\lambda_{\rho}}{ \lambda_{\sigma}}
&=\inf_{\sigma \in \Sigma } \kl{\lambda_{\rho}}{ \lambda_{\sigma}} \leq \inf_{\sigma \in  \cJ\left(\cE_{\rho},\Sigma\right)} \kl{\lambda_{\rho}}{ \lambda_{\sigma}},
 \notag
\end{align}
and the quantities at left and right extremes are equal by \eqref{eq:mlpexctexp}. In the above equation, only the penultimate equality is non-trivial, i.e.,
\begin{align}
    \inf_{\sigma \in \Sigma(\cE_{\rho})} \kl{\lambda_{\rho}}{ \lambda_{\sigma}}
=\inf_{\sigma \in \Sigma } \kl{\lambda_{\rho}}{ \lambda_{\sigma}}. \notag
\end{align}
The  $\geq $ implication is straightforward as $\Sigma(\cE_{\rho}) \subseteq \Sigma$, while the $\leq $ implication follows from the fact that for any $\sigma \in \Sigma $, there exists another $\sigma' \in \Sigma(\cE_{\rho})$ with the same eigenvalues $\lambda_{\sigma}$ since $ \Sigma$ is unitarily invariant.
\subsection{Proof of Proposition \ref{Prop:devmlpred}}\label{Sec:Prop:devmlpred-proof}
We have 
\begin{align}
  2 \big(\qrel{\tilde{\rho}}{ \tilde{\sigma}^{\star}}-\qrel{\rho}{\sigma^{\star}}\big) &\stackrel{(a)}{\geq}\norm{\tilde{\rho}- \tilde{\sigma}^{\star}}_1^2 -2\epsilon \notag \\
    & \stackrel{(b)}{\geq}0.5 \norm{ \tilde{\sigma}^{\star}-\sigma^{\star}}_1^2- \norm{\tilde{\rho}-\sigma^{\star}}_1^2-2\epsilon \notag \\
    & \stackrel{(c)}{\geq} 0.5 \norm{ \tilde{\sigma}^{\star}-\sigma^{\star}}_1^2-2\norm{\tilde{\rho}-\rho}_1^2-2 \norm{\rho-\sigma^{\star}}_1^2-2\epsilon \notag \\
& \stackrel{(d)}{\geq}  0.5 \norm{ \tilde{\sigma}^{\star}-\sigma^{\star}}_1^2 -2 \norm{\tilde{\rho}-\rho}_1^2-6\epsilon, \notag
\end{align}
where $(a)$ and $(d)$ are due to $\mathsf{D}(\rho \| \sigma^{\star}) \leq \epsilon$ and quantum Pinsker's inequality: $2 \qrel{\rho}{\sigma} \geq \norm{\rho-\sigma}_1^2$ for $\rho,\sigma \in \cS(\HH)$, while $(b)$ and $(c)$ follows by applying $(a+b)^2 \leq 2 (a^2+b^2) $. 
Hence, \eqref{eq:trdistdenop1} follows since
\begin{align}
\norm{\tilde{\sigma}^{\star}-\sigma^{\star}}_1^2 &\leq 4 \left(\mathsf{D}\big(\tilde{\rho} \| \tilde{\sigma}^{\star}\big)-\mathsf{D}\big(\rho\| \sigma^{\star}\big) \right)  +4 \norm{\tilde{\rho}-\rho}_1^2+12\epsilon \notag \\
&\leq 4 \left(\mathsf{D}\big(\tilde{\rho} \| \sigma^{\star}\big)-\mathsf{D}\big(\rho\| \sigma^{\star}\big) \right)  +4 \norm{\tilde{\rho}-\rho}_1^2+12\epsilon. \label{eq:ubndtvd}  
\end{align}  
Equation \eqref{eq:trdistdenop2} is a simple consequence of triangle inequality for trace distance and quantum Pinsker's inequality along with $\mathsf{D}(\rho \| \sigma^{\star}) \leq \epsilon$:
\begin{align}
    \norm{\tilde{\sigma}^{\star}-\rho}_1 \leq  \norm{\tilde{\sigma}^{\star}-\sigma^{\star}}_1 + \norm{\sigma^{\star}-\rho}_1 \leq \norm{\tilde{\sigma}^{\star}-\sigma^{\star}}_1+ \sqrt{2\epsilon}. \notag
\end{align} 

\medskip
Next, consider the case of unitary invariant $\Sigma$ such that $\cJ\left(\cE_{\rho},\Sigma\right) \subseteq \Sigma $. Applying \eqref{eq:mlpexctexp},  we have 
\begin{align}
  \mathsf{D}\big(\rho\| \sigma^{\star}\big)=\min_{\sigma \in \Sigma}   \mathsf{D}\big(\rho\| \sigma\big)= \min_{\sigma \in  \cJ\left(\cE_{\rho},\Sigma\right)} \kl{\lambda_{\rho}}{\lambda_{\sigma}}=\kl{\lambda_{\rho}}{\lambda_{\sigma^{\star}}}. \label{eq:qrelkl1} 
\end{align} 
On the other hand
\begin{align}
    \mathsf{D}\big(\tilde{\rho} \| \tilde{\sigma}^{\star}\big)=\min_{\sigma \in \Sigma}    \mathsf{D}\big(\tilde{\rho} \| \sigma\big) = \min_{\sigma \in \Sigma}    \kl{\lambda_{\tilde{\rho}}}{\lambda_{\sigma}} \leq \kl{\lambda_{\tilde{\rho}}}{\lambda_{\sigma^{\star}}}, \label{eq:qrelkl2}
\end{align}
where the second equality above is due to the last statement in Proposition \ref{prop:QMLPrel} which holds since  $\Sigma$ is unitarily invariant, compact, and satisfies $\cJ\left(\cE_{\rho},\Sigma\right) \subseteq \Sigma $. Combining \eqref{eq:ubndtvd}, \eqref{eq:qrelkl1} and \eqref{eq:qrelkl2} results in the final claim. 
\subsection{Proof of Theorem \ref{Thm:convergence-MLpred}} \label{Sec:Thm:convergence-MLpredKL-proof}
First, we prove Part $(i)$. 
We  require two concentration results. 
\begin{theorem}[Matrix Hoeffding and Bernstein Inequalities, see e.g., \citep{Vershynin2018HighDimensionalProbability}]\label{Thm:matbernhoeff}
    Let $\{H_i\}_{i=1}^n$ be fixed Hermitian $d \times d$ matrices. Let $\{\varepsilon_i\}_{i=1}^n$ be i.i.d. Rademacher  variables. Then
    \begin{subequations}
        \begin{align}
       & \PP\left(\norm{\sum_{i=1}^n \varepsilon_i H_i } \geq t\right) \leq 2d e^{-\frac{t^2}{2V_n^2}},~\forall ~t \geq 0, \label{eq:concmathoeffding} \\
       & \EE\left[\norm{\sum_{i=1}^n \varepsilon_i H_i }\right] \leq \left(\sqrt{2\log (2d)}+\sqrt{\pi/2}\right) V_n,\label{eq:expbndmathoeff} 
    \end{align}   
    \end{subequations}
    where $V_n^2\coloneqq \norm{\sum_{i=1}^n H_i^2}$.
   If $\{H_i\}_{i=1}^n$ are random independent zero-mean $d \times d $ Hermitian matrices such that $\norm{H_i} \leq M$ almost surely, then
    \begin{align}
                & \PP\left(\norm{\sum_{i=1}^n H_i } \geq t\right) \leq 2d e^{-\left(\frac{t^2}{4\bar V_n^2} \wedge \frac{3t}{4 M}\right)},~\forall ~t \geq 0, \label{eq:concmatbernstein} 
    \end{align}
    where $\bar V_n^2 \coloneqq  \norm{\sum_{i=1}^n \EE\big[H_i^2\big]}$.
\end{theorem}

\medskip 

We proceed with the proof of \eqref{eq:conv-rate} when $\epsilon>0$. Note that since $\Sigma$ is a compact  set and $\supp(\Sigma)=\supp(\rho_p)=\HH_d$, $\sigma_p^{\star}>0$ exists.    
From \eqref{eq:trdistdenop}, we have using $(a+b)^2 \leq 2(a^2+b^2)$ that
\begin{align}
     \norm{\sigma_n^{\star}(X^n)-\rho_p}_1^2 &\leq 8\left(\qrel{\rho_n(X^n)}{\sigma_p^{\star}}-\qrel{\rho_p}{\sigma_p^{\star}}\right) +8 \norm{\rho_n(X^n)-\rho_p}_1^2+28\epsilon. \label{eq:tvsqrbnd}
\end{align}
To control the first term in \eqref{eq:tvsqrbnd}, we will use the variational expression for quantum relative entropy given in \eqref{eq:varexpqrelent}. When $\rho,\sigma>0$, the supremum in \eqref{eq:varexpqrelent} is achieved by
\begin{align}
H^{\star}(\rho,\sigma) \coloneqq \log \rho -\log \sigma. \notag 
\end{align}
Since $\rho_p>0$ and $\sigma_p^{\star}>0$, it follows that 
\begin{align}
  \qrel{\rho_p}{\sigma_p^{\star}}  =\sup_{H} \tr{\rho_p H}-\tr{e^{H+\log \sigma_p^{\star}}}+1, \label{eq:varexprel3} 
\end{align}
where the supremum may be restricted to $H$ satisfying
\begin{align}
\norm{H}\leq \norm{H^{\star}(\rho_p,\sigma_p^{\star})}_{\infty}
\leq T(\rho_p,\sigma_p^{\star}). \notag
\end{align}
Indeed, if $T=T(\rho_p,\sigma_p^{\star})$, then $e^{-T}\sigma_p^{\star}\leq\rho_p\leq e^T\sigma_p^{\star}$. Operator monotonicity of the logarithm gives $-TI\leq \log\rho_p-\log\sigma_p^{\star}\leq TI$, and hence the stated norm bound. 
Similarly
\begin{align}
    \qrel{\rho_n(X^n)}{\sigma_p^{\star}} =\sup_{H} \tr{\rho_n(X^n) H}-\tr{e^{H+\log \sigma_p^{\star}}}+1,
\end{align}
where the supremum may be restricted to $H$ satisfying
\begin{align}
\norm{H}\leq \norm{H^{\star}(\rho_n(X^n),\sigma_p^{\star})}_{\infty}
\leq T(\rho_n(X^n),\sigma_p^{\star})
\leq \log\norm{\sigma_p^{\star-1}}\vee\log(dn). \notag
\end{align}
The last inequality follows from $\norm{\rho_n(X^n)}_{\infty}\vee\norm{\sigma_p^{\star}}_{\infty}\leq1$, $\rho_n(X^n)\geq I/(dn)$, and  
\begin{align}
T(\rho_n(X^n),\sigma_p^{\star})&=\log \left(\max_{\ket{v}: \norm{v}_2=1} \frac{\bra{v} \rho_n(X^n)\ket{v}}{\bra{v} \sigma_p^{\star}\ket{v}} \vee \max_{\ket{v}: \norm{v}_2=1} \frac{\bra{v} \sigma_p^{\star}\ket{v}}{\bra{v} \rho_n(X^n)\ket{v}}\right), \notag \\
   \max_{\ket{v}: \norm{v}_2=1} \frac{\bra{v} \rho_n(X^n)\ket{v}}{\bra{v} \sigma_p^{\star}\ket{v}} & \leq \frac{1}{\min_{\ket{v}: \norm{v}_2=1}\bra{v} \sigma_p^{\star}\ket{v}} \leq \frac{1}{\min_{z \in \cZ} \lambda_{\sigma_p^{\star}}(z)}, \notag \\
    \max_{\ket{v}: \norm{v}_2=1} \frac{\bra{v} \sigma_p^{\star}\ket{v}}{\bra{v} \rho_n(X^n)\ket{v}} &\leq  \frac{1}{\min_{\ket{v}: \norm{v}_2=1}\bra{v} \rho_n(X^n)\ket{v}} \leq nd. \notag
\end{align}
Hence, with 
$b_n$ as in \eqref{eq:bndfuncclls}, 
we can upper bound the first term in the right hand side (RHS) above as follows:
\begin{align}
&\abs{\kl{\rho_n(X^n)}{\sigma_p^{\star}}-\kl{\rho_p}{\sigma_p^{\star}}} \notag \\
&=  \abs{\sup_{H:\norm{H} \leq b_n } \tr{\rho_n(X^n) H}-\tr{e^{H+\log \sigma_p^{\star}}} -\sup_{H:\norm{H} \leq b_n } \tr{\rho_p H}-\tr{e^{H+\log \sigma_p^{\star}}} }\notag \\
&\stackrel{(a)}{\leq} \sup_{H:\norm{H} \leq b_n }\abs{\tr{\rho_n(X^n) H}-\tr{\rho_p H}}. \notag \\
&\stackrel{(b)}{\leq} \left(1-\frac{1}{n}\right)\sup_{H:\norm{H} \leq b_n }\abs{\tr{\hat \rho_n(X^n) H}-\tr{\rho_p H}}+\sup_{H:\norm{H} \leq b_n } \frac{1}{n }\abs{\tr{\pi_d H}}+\frac{1}{n }\abs{\tr{\rho_p H}}. \notag \\
& \stackrel{(c)}{\leq}\norm{\hat \rho_n(X^n) -\rho_p}_1 b_n  + \frac{2b_n}{n}, \notag 
\end{align}
where 
\begin{enumerate}[(a)]
    \item follows from 
\begin{align}
\abs{\sup_{t\in\mathcal T} f(t)-\sup_{t\in\mathcal T} g(t)} \leq \sup_{t\in\mathcal T}\abs{f(t)-g(t)}, \label{eq:bnddiffsup}    
\end{align}
 for any functions $f,g$ on  $\mathcal T$ when at least  $\sup_{t\in\mathcal T} g(t) $ or $\sup_{t\in\mathcal T} f(t) $ is finite;
\item follows by applying definition of $\rho_n(X^n)$ and triangle inequality;
\item is via H\"older's inequality. 
\end{enumerate}
Substituting this in \eqref{eq:tvsqrbnd} and using $\norm{\rho_n(X^n)-\rho_p}_1^2 \leq 2\big(\norm{\hat \rho_n(X^n)-\rho_p}_1^2+(4/n^2)\big)$, we obtain
\begin{align}
 \norm{\sigma_n^{\star}(X^n)-\rho_p}_1^2 &\leq 8b_n\norm{\hat \rho_n(X^n) -\rho_p}_1   + \frac{16 b_n}{n} +16 \norm{\hat \rho_n(X^n)-\rho_p}_1^2+\frac{64}{n^2}+28\epsilon.\label{eq:bndtvdistsq}    
\end{align}
Taking expectations, we can bound the first term in the RHS as 
    \begin{align}
       \EE\big[\norm{\hat \rho_n(X^n)-\rho_p}_1\big] 
       & \stackrel{(a)}{\leq}  d ~\EE \left[\norm{\frac{1}{n} \sum_{i=1}^n \big(\ket{\varphi(X_i)} \bra{\varphi(X_i)}-\EE_{P}\big[\ket{\varphi(X)} \bra{\varphi(X)}\big]\big)}\right] \notag \\
       & \stackrel{(b)}{\leq} 2d~ \EE \left[\norm{\frac{1}{n} \sum_{i=1}^n \varepsilon_i\ket{\varphi(X_i)} \bra{\varphi(X_i)}}\right] \notag \\
         & \stackrel{(c)}{\leq} 2d \left(\sqrt{2\log (2d)}+\sqrt{\pi/2}\right) n^{-\frac 12}, \label{eq:bndtrdistrho}
    \end{align}
    where 
    \begin{enumerate}[(a)]
    \item is due to $\norm{H}_1 \leq d \norm{H}$ for a self-adjoint $H$;
    \item is due to the standard  symmetrization inequality (see e.g. \citep[Lemma 6.4.2]{Vershynin2018HighDimensionalProbability});
    \item follows by an application of \eqref{eq:expbndmathoeff}  with $H_i=\ket{\varphi(x_i)}\bra{\varphi(x_i)}/n$ for  a given $X^n=x^n$, and noting that 
    \begin{align}
        V_n^2 \mspace{-2 mu}&\coloneqq \mspace{-2 mu} \frac{1}{n^2}\norm{\sum_{i=1}^n \ket{\varphi(x_i)}\bra{\varphi(x_i)}\ket{\varphi(x_i)}\bra{\varphi(x_i)}} \mspace{-2 mu} = \mspace{-2 mu}\frac{1}{n^2}\norm{\sum_{i=1}^n \innp{\varphi(x_i)}{\varphi(x_i)}\ket{\varphi(x_i)}\bra{\varphi(x_i)}} 
        \mspace{-2 mu}\leq \mspace{-2 mu} \frac{1}{n}, \notag 
    \end{align}
     where  the final inequality uses 
    $\innp{\varphi(x)}{\varphi(x)} =1$ for all $x \in \cX$, $\norm{\ket{\varphi(x_i)}\bra{\varphi(x_i)}} \leq 1$ and triangle inequality for operator norm.
    \end{enumerate}
        For the second term in \eqref{eq:bndtvdistsq}, using $0.5 \norm{\hat \rho_n(X^n)-\rho_p}_1 \leq 1$ and $z^2 \leq z$ for $0 \leq z \leq 1$,  \eqref{eq:bndtrdistrho} implies 
    \begin{align}
       \EE \big[\norm{\hat \rho_n(X^n)-\rho_p}_1^2\big] \leq   2\EE \big[\norm{\hat \rho_n(X^n)-\rho_p}_1\big]  \leq 4d \left(\sqrt{2\log (2d)}+\sqrt{\pi/2}\right) n^{-\frac 12}. \label{eq:tvsqbnd}
    \end{align}
        Substituting  \eqref{eq:bndtrdistrho} and \eqref{eq:tvsqbnd} in \eqref{eq:bndtvdistsq}, we obtain \eqref{eq:conv-rate} when $\epsilon>0$.

\medskip

 Before proving \eqref{eq:conv-rate} when $\epsilon=0$, we first establish \eqref{eq:devineqmlpred}. We  have from \eqref{eq:bndtvdistsq} that for $t \geq 0$:
    \begin{align}
&\PP\left(\norm{\sigma_n^{\star}(X^n)-\rho_p}_1^2 \geq   \frac{16 b_n}{n} +\frac{64}{n^2}+28\epsilon+8d (b_n+4)t\right)\notag \\ 
    & \leq \PP\left(\norm{\hat \rho_n(X^n)-\rho_p}_1 \geq d t \right)+\PP\left(\norm{\hat \rho_n(X^n)-\rho_p}_1^2 \geq 2dt\right).\label{eq:devineqspl} 
    \end{align}  
    For the first term in \eqref{eq:devineqspl}, an application of  \eqref{eq:concmatbernstein} with $H_i= \ket{\varphi(X_i)} \bra{\varphi(X_i)}-\EE_P\big[\ket{\varphi(X)} \bra{\varphi(X)}\big]$ yields
    \begin{align}
    \PP\left(\norm{\hat \rho_n(X^n)-\rho_p}_1 \geq dt \right) \leq \PP\left(\norm{\hat \rho_n(X^n)-\rho_p} \geq t \right) \leq 2d e^{-\left(\frac{nt^2}{4} \wedge \frac{3nt}{4}\right)},   \label{eq:trnormdev}
    \end{align}
    where we used $\norm{ H_i} \leq 1$ and  $\norm{\sum_{i=1}^n H_i^2} \leq \sum_{i=1}^n \norm{ H_i^2} \leq \sum_{i=1}^n \norm{ H_i} \leq n $.  
   For the second term in \eqref{eq:devineqspl}, we  have via \eqref{eq:concmatbernstein} that
    \begin{align}
    \PP\left(\norm{\hat \rho_n(X^n)-\rho}_1^2 \geq 2dt \right)  &\leq \PP\left(\frac{\norm{\hat \rho_n(X^n)-\rho}_1^2}{4} \geq \frac{dt}{2} \right) \notag \\
    &\leq \PP\left(\frac{\norm{\hat \rho_n(X^n)-\rho}_1}{2} \geq \frac{dt}{2} \right) \notag \\
    &\leq \PP\left(\norm{\hat \rho_n(X^n)-\rho} \geq t\right) \notag \\
    &\leq 2d e^{-\left(\frac{n t^2}{4} \wedge \frac{3nt}{4}\right)}.\notag
    \end{align}
  Substituting this and \eqref{eq:trnormdev} in \eqref{eq:devineqspl}  yields \eqref{eq:devineqmlpred}. 

\medskip

Next, we show \eqref{eq:conv-rate} when $\epsilon=0$,  i.e., $\sigma_p^{\star}=\rho_p$. Then, \eqref{eq:tvsqrbnd} yields that
\begin{align}
     \norm{\sigma_n^{\star}(X^n)-\rho_p}_1^2 &\leq 8\qrel{\rho_n(X^n)}{\rho_p} +8 \norm{\rho_n(X^n)-\rho_p}_1^2. \label{eq:tvsqrbndmatch}
\end{align}
Applying a second order Taylor's expansion as in \citep[Equation 41]{SB-IT-2025}, we have
\begin{align}
   \qrel{\rho_n(X^n)}{\rho_p} &= 2\operatorname{Tr}\mspace{-2 mu}\bigg[\int_{0}^1 (1-t)   (\rho_n(X^n)-\rho_p) \int_{0}^{\infty}\mspace{-2 mu} u_1(\rho_n(X^n),\rho_p,\tau,t)   d\tau \bigg] dt  \notag \\& \qquad -2\int_{0}^1 (1-t)\operatorname{Tr}\bigg[\big((1-t)\rho_p+t\rho_n(X^n)\big)  
     \int_{0}^{\infty} u_2(\rho_n(X^n),\rho_p,\tau,t) d\tau \bigg] dt, \label{eq:qreltayexp}
\end{align}
 where 
\begin{subequations}\label{eq:intermtermstayexp}
  \begin{align}
v(\rho_n(X^n),\rho_p,\tau,t)&:=\big(\tau I+(1-t)\rho_p+t\rho_n(X^n)\big)^{-1}, \\
u_1(\rho_n(X^n),\rho_p,\tau,t)&:=v(\rho_n(X^n),\rho_p,\tau,t)(\rho_n(X^n)-\rho_p)v(\rho_n(X^n),\rho_p,\tau,t), \\ 
u_2(\rho_n(X^n),\rho_p,\tau,t) &=u_1(\rho_n(X^n),\rho_p,\tau,t)(\rho_n(X^n)-\rho_p)v(\rho_n(X^n),\rho_p,\tau,t). 
\end{align}  
\end{subequations}
Using H\"{o}lders inequality and sub-multiplicavity of Schatten norms, we obtain
\begin{align}
&\operatorname{Tr}\bigg[\int_{0}^1 (1-t)   (\rho_n(X^n)-\rho_p) \int_{0}^{\infty} u_1(\rho_n(X^n),\rho_p,\tau,t)   d\tau \bigg] dt  \notag \\
& \leq \norm{\rho_n(X^n)-\rho_p}_1 \int_{0}^1 (1-t)\bigg[\int_{0}^{\infty} \norm{u_1(\rho_n(X^n),\rho_p,\tau,t)}   d\tau \bigg] dt \notag \\
& \leq \norm{\rho_n(X^n)-\rho_p}_1 \norm{\rho_n(X^n)-\rho_p}\int_{0}^1 (1-t)\int_{0}^{\infty}\norm{\big(\tau I+(1-t)\rho_p+t\rho_n(X^n)\big)^{-1}}^2   d\tau ~ dt \notag \\
& \leq \norm{\rho_n(X^n)-\rho_p}_1 \norm{\rho_n(X^n)-\rho_p}\int_{0}^1 (1-t)\int_{0}^{\infty}\norm{\big(\tau I+(1-t)\rho_p\big)^{-1}}^2   d\tau ~ dt \notag \\
& \leq \norm{\rho_n(X^n)-\rho_p}_1 \norm{\rho_n(X^n)-\rho_p}\int_{0}^1 (1-t)\int_{0}^{\infty} (\tau +(1-t) \lambda_{\rho_p}^{\min})^{-2}  d\tau ~ dt \notag \\
& \leq \norm{\rho_n(X^n)-\rho_p}_1 \norm{\rho_n(X^n)-\rho_p} \norm{\rho_p^{-1}}. \label{eq:bndtay1} 
\end{align}
Similarly, we have
\begin{align}
&\int_{0}^1 (1-t)\operatorname{Tr}\bigg[\big((1-t)\rho_p+t\rho_n(X^n)\big)  
     \int_{0}^{\infty} u_2(\rho_n(X^n),\rho_p,\tau,t) d\tau \bigg]    \notag \\
     & \stackrel{(a)}{\leq} \norm{\rho_n(X^n)-\rho_p}_1 \norm{\rho_n(X^n)-\rho_p} \int_{0}^1 (1-t)\bigg[\int_{0}^{\infty}\norm{\big(\tau I+(1-t)\rho_p\big)^{-1}}^2   d\tau \bigg]~ dt \notag \\
     & \leq \norm{\rho_n(X^n)-\rho_p}_1 \norm{\rho_n(X^n)-\rho_p}\norm{\rho_p^{-1}}, \label{eq:bndtay2}
\end{align}
where $\norm{(1-t)\rho_p+t\rho_n(X^n)}  \big\|\big(\tau I+(1-t)\rho_p+t\rho_n(X^n)\big)^{-1}\big\| \leq 1$. Substituting  \eqref{eq:bndtay1} and \eqref{eq:bndtay2} in \eqref{eq:qreltayexp}, we obtain
\begin{align}
    \qrel{\rho_n(X^n)}{\rho_p} \leq 4 \norm{\rho_n(X^n)-\rho_p}_1 \norm{\rho_n(X^n)-\rho_p} \norm{\rho_p^{-1}}.\notag
\end{align}
Substituting the above inequality in \eqref{eq:tvsqrbndmatch}, we obtain
\begin{align}
     \norm{\sigma_n^{\star}(X^n)-\rho_p}_1^2 &\leq 32 \norm{\rho_n(X^n)-\rho_p}_1 \norm{\rho_n(X^n)-\rho_p} \norm{\rho_p^{-1}} +8 \norm{\rho_n(X^n)-\rho_p}_1^2 \notag \\
     & \leq 32 d \norm{\rho_n(X^n)-\rho_p}^2  \norm{\rho_p^{-1}} +8 \norm{\rho_n(X^n)-\rho_p}_1^2 \notag \\
     & \leq (32 d \norm{\rho_p^{-1}}+ 8 d^2)\norm{\rho_n(X^n)-\rho_p}^2, \label{eq:tvbnd-matched}
\end{align}
where the final inequality uses $\norm{\rho_n(X^n)-\rho_p}_1 \leq d \norm{\rho_n(X^n)-\rho_p}$. 
Taking expectations, we obtain
\begin{align}
    \EE \left[\norm{\sigma_n^{\star}(X^n)-\rho_p}_1^2\right] \leq (32 d \norm{\rho_p^{-1}}+ 8 d^2) \EE \left[\norm{\rho_n(X^n)-\rho_p}^2 \right]. \label{eq:expqmlperr} 
\end{align}
We will show that the last expectation is $O(1/n)$. An application of  \eqref{eq:concmatbernstein} with $H_i= \ket{\varphi(X_i)} \bra{\varphi(X_i)}-\EE_P\big[\ket{\varphi(X)} \bra{\varphi(X)}\big]$ yields
    \begin{align}
   \PP\left(\norm{\hat \rho_n(X^n)-\rho_p} \geq t \right) \leq 2d e^{-\left(\frac{nt^2}{4} \wedge \frac{3nt}{4}\right)},  
    \end{align}
    where we used $\norm{ H_i} \leq 1$ and  $\norm{\sum_{i=1}^n H_i^2} \leq \sum_{i=1}^n \norm{ H_i^2} \leq \sum_{i=1}^n \norm{ H_i} \leq n $.   Thus, we have
\begin{align}
\EE \left[\norm{\hat \rho_n(X^n)-\rho_p}^2 \right] &= \int_{0}^{\infty} \PP\left(\norm{\hat \rho_n(X^n)-\rho_p} \geq \sqrt{t}\right) dt\notag \\
&\leq 2d \int_{0}^{\infty}   e^{-\left(\frac{nt}{4} \wedge \frac{3n\sqrt{t}}{4}\right)}dt \notag \\
&\leq 2d \int_{0}^{9}   e^{-\frac{nt}{4} } dt+2d \int_{9}^{\infty}   e^{-\frac{3n\sqrt{t}}{4}} dt \notag \\
&\leq\frac{8d}{n} + \frac{16d e^{-\frac{9n}{4}}}{3n}\left(3+\frac{4}{3n}\right). \label{eq:bndexpempsum}
\end{align}
Substituting \eqref{eq:bndexpempsum} in \eqref{eq:expqmlperr} and using
$\norm{\rho_n(X^n)-\rho_p}_1^2 \leq 2\big(\norm{\hat \rho_n(X^n)-\rho_p}_1^2+(4/n^2)\big)$, we obtain
\begin{align}
    \EE \left[\norm{\sigma_n^{\star}(X^n)-\rho_p}_1^2\right] &\leq (32 d \norm{\rho_p^{-1}}+ 8 d^2) \EE \left[\norm{\rho_n(X^n)-\rho_p}^2 \right]  \notag \\
    & \leq (32 d \norm{\rho_p^{-1}}+ 8 d^2) \left(\frac{8}{n^2} + \frac{16d}{n} + \frac{32d e^{-\frac{3n}{4}}}{3n}\left(1+\frac{4}{3n}\right)\right) \notag \\
    & \leq (32 d \norm{\rho_p^{-1}}+ 8 d^2) \left(\frac{8}{n^2} + \frac{28d}{n}\right), \notag
\end{align}
thus proving \eqref{eq:conv-rate} when $\epsilon=0$.

\medskip

The proof of Part $(i)$ is completed by noting that \eqref{eq:conc-ineq} follows using \eqref{eq:tvbnd-matched} and  $\norm{\rho_n(X^n)-\rho_p}_1^2 \leq 2\big(\norm{\hat \rho_n(X^n)-\rho_p}_1^2+(4/n^2)\big)$ as follows:
  \begin{align}
\PP\left(\norm{\sigma_n^{\star}(X^n)-\rho_p}_1^2 \geq   (32 d \norm{\rho_p^{-1}}+ 8 d^2)\left(\frac{8}{n^2}+2t\right) \right) &\leq \PP\left(\norm{\hat \rho_n(X^n)-\rho_p}^2 \geq t\right)  \notag \\
&\leq 2d e^{-\left(\frac{nt}{4} \wedge \frac{3n\sqrt{t}}{4}\right)}. \notag
\end{align}

\medskip

Next, we prove Part $(ii)$ starting with the proof of \eqref{eq:experrorqrel} when $\epsilon>0$.  We have 
\begin{align}
  &  \qrel{\rho_p}{\hat{\sigma}_n^{\star}(X^n)} \notag \\
  &=  \qrel{\rho_p}{\hat{\sigma}_n^{\star}(X^n)}- \qrel{\rho_n(X^n)}{\hat{\sigma}_n^{\star}(X^n)}+\qrel{\rho_n(X^n)}{\hat{\sigma}_n^{\star}(X^n)} \notag \\
    & \stackrel{(a)}{\leq} \qrel{\rho_p}{\hat{\sigma}_n^{\star}(X^n)}- \qrel{\rho_n(X^n)}{\hat{\sigma}_n^{\star}(X^n)}+\qrel{\rho_n(X^n)}{\sigma_n^{\star}(X^n)} +\frac{1}{n}\qrel{\rho_n(X^n)}{\pi_d} \notag \\
       & \stackrel{(b)}{\leq} \qrel{\rho_p}{\hat{\sigma}_n^{\star}(X^n)}- \qrel{\rho_n(X^n)}{\hat{\sigma}_n^{\star}(X^n)}+\qrel{\rho_n(X^n)}{\sigma_n^{\star}(X^n)} +\frac{\log d}{n}, \label{eq:empklrisk}
\end{align}
where $(a)$ follows by convexity of quantum relative entropy and $(b)$ follows since $\qrel{\rho_n(X^n)}{\pi_d}\leq \log d$. 
Taking expectations with respect to $X^n \sim P^{\otimes n}$, we obtain under the assumption in  \eqref{eq:klassump} that
\begin{align}
 \EE \left[\qrel{\rho_p}{\hat{\sigma}_n^{\star}(X^n)}\right] \leq   \EE \left[  \qrel{\rho_p}{\hat{\sigma}_n^{\star}(X^n)}- \qrel{\rho_n(X^n)}{\hat{\sigma}_n^{\star}(X^n)}\right]+\frac{\log d}{n}+ \epsilon. \label{eq:qrelrisk}
\end{align}
In the above, we used $\EE\left[\qrel{\rho_n(X^n)}{\sigma_n^{\star}(X^n)}\right] \leq \epsilon$, however, if $\EE\left[\qrel{\rho_n(X^n)}{\hat{\sigma}_n^{\star}(X^n)}\right] \leq \epsilon$ holds, then the same bound in \eqref{eq:qrelrisk} is valid by skipping the steps $\stackrel{(a)}{\leq}$ and $\stackrel{(b)}{\leq}$ leading to \eqref{eq:empklrisk} and using this condition instead. Hence, \eqref{eq:qrelrisk} is valid when \eqref{eq:klassump} holds.

Since $\rho_p>0$ and $\hat{\sigma}_n^{\star}(x^n)>0$ for every $x^n \in \cX^n$, it follows that 
\begin{align}
  \qrel{\rho_p}{\hat{\sigma}_n^{\star}(X^n)}  =\sup_{H} \tr{\rho_p H}-\tr{e^{H+\log \hat{\sigma}_n^{\star}(X^n)}}+1, \label{eq:varexprel1} 
\end{align}
where the supremum is over all  $H$ such that
\begin{align}
\norm{H} \leq \norm{H^{\star}\big(\rho_p,\hat{\sigma}_n^{\star}(X^n)\big)} \leq \log (dn) \vee \log \norm{\rho_p^{-1}}.\notag 
\end{align}
Similarly,
\begin{align}
    \qrel{\rho_n(X^n)}{\hat{\sigma}_n^{\star}(X^n)}=\sup_{H} \tr{\rho_n(X^n)H}-\tr{e^{H+\log \hat{\sigma}_n^{\star}(X^n)}}+1, \label{eq:varexprel2}
\end{align}
where the supremum is over all $H$ such that 
\begin{align}
\norm{H} \leq \norm{H^{\star}\big(\rho_n(X^n),\hat{\sigma}_n^{\star}(X^n)\big)} \leq \log(dn).\notag
\end{align}
Hence, we can restrict the suprema in \eqref{eq:varexprel1} and \eqref{eq:varexprel2} over $H$ such that $\norm{H} \leq \bar{b}_n \coloneqq \log (dn) \vee \log \norm{\rho_p^{-1}}$. Then,  we obtain
\begin{align}
 & \abs{\qrel{\rho_p}{\hat{\sigma}_n^{\star}(X^n)}- \qrel{\rho_n(X^n)}{\hat{\sigma}_n^{\star}(X^n)} }\notag \\
 &\stackrel{(a)}{\leq} \sup_{H: \norm{H} \leq \bar{b}_n} \abs{\tr{\big(\rho_n(X^n)-\rho_p\big)H }}  \notag \\
  & \stackrel{(b)}{\leq} \sup_{H: \norm{H} \leq \bar{b}_n} \frac{\abs{\tr{H}}}{dn}+\frac{\abs{\tr{\rho_p H}}}{n} +\abs{\tr{\big(\hat{\rho}_n(X^n)-\rho_p\big)H }} \notag \\
  &\stackrel{(c)}{\leq}  \frac{2\bar{b}_n}{n}+ \sup_{H: \norm{H}\leq \bar{b}_n} \abs{\tr{\big(\hat{\rho}_n(X^n)-\rho_p\big)H }} \notag \\
  &\stackrel{(d)}{\leq}  \frac{2\bar{b}_n}{n}+ d \bar{b}_n \norm{\hat{\rho}_n(X^n)-\rho_p},\notag 
\end{align}
where 
\begin{enumerate}[(a)]
    \item follows by applying \eqref{eq:bnddiffsup};
    \item follows by using the definition of $\rho_n(X^n)$;
    \item is because $\abs{\tr{H}} \leq d \bar{b}_n$ and by H\"{o}lder's inequality,  $\abs{\tr{\rho_p H}} \leq \norm{\rho_p}_1\norm{H} \leq \bar{b}_n$;
    \item is again via  H\"{o}lder's inequality and $\norm{H}_1 \leq d \bar{b}_n$.
\end{enumerate}
Hence, applying \eqref{eq:expbndmathoeff} in Theorem \ref{Thm:matbernhoeff} similar to \eqref{eq:bndtrdistrho}, we obtain
\begin{align}
   \EE \left[  \abs{\qrel{\rho_p}{\hat \sigma_n^{\star}(X^n)}- \qrel{\rho_n(X^n)}{\hat \sigma_n^{\star}(X^n)}}\right] &\leq  \frac{2\bar{b}_n}{n}+ d \bar{b}_n \EE\left[\norm{\hat{\rho}_n(X^n)-\rho_p}\right] \notag \\
   & \leq 2\bar{b}_n n^{-1}+ d \bar{b}_n\left(\sqrt{2\log (2d)}+\sqrt{\pi/2}\right) n^{-\frac 12}. \notag 
\end{align}
The claim in \eqref{eq:experrorqrel} for the case $\epsilon>0$ then follows by substituting this in \eqref{eq:qrelrisk}.

To obtain the bound in \eqref{eq:experrorqrel} for the case $\epsilon=0$, we note that \eqref{eq:klassump} and non-negativity of quantum relative entropy implies that  either   $\sigma_n^{\star}(X^n)=\rho_n(X^n)$ or $\hat{\sigma}_n^{\star}(X^n)=\rho_n(X^n)$ holds almost surely. In both cases, we have using convexity of quantum relative entropy and the definitions in \eqref{eq:empMLdist}-\eqref{eq:empMLpred} that
\begin{align}
 \qrel{\rho_p}{\hat{\sigma}_n^{\star}(X^n)} \leq  \qrel{\rho_p}{\rho_n(X^n)} +\frac{\log d}{n}.\label{eq:bndqrelmain}
\end{align}
Set  $\kappa_p=\norm{\rho_p^{-1}}$, $\lambda_p=\kappa_p^{-1}$ and, for $n>2\kappa_p$, $\delta_n=\lambda_p/2-1/n$. Consider the event
\begin{align}
\mathcal G_n\coloneqq\{\norm{\hat\rho_n(X^n)-\rho_p}\leq\delta_n\}. \notag
\end{align}
Under the event  $\mathcal G_n$, using $\norm{\pi_d-\rho_p}\leq1$,
\begin{align}
\norm{\rho_n(X^n)-\rho_p}\leq\norm{\hat\rho_n(X^n)-\rho_p}+\frac1n\leq\frac{\lambda_p}{2}, \notag
\end{align}
and hence $\rho_n(X^n)\geq\rho_p/2$. Therefore $\rho_n(X^n)^{-1}\leq2\rho_p^{-1}$.

Next, we recall that quantum relative entropy can be upper bounded by a version of quantum chi-squared divergence \citep[Theorem 8]{Temme-2010}:
\begin{align}
\qrel{\rho}{\sigma} \leq \chi^2(\rho \|\sigma), \label{eq:chisquppbnd}
\end{align}
where 
\begin{align} \chi^2(\rho \|\sigma)\coloneqq \begin{cases}
    \tr{(\rho-\sigma)\sigma^{-1}(\rho-\sigma)}, & \mbox{ if } \rho \ll \sigma, \\
    \infty, & \mbox{ otherwise}.
\end{cases} \notag
\end{align}
From this, we obtain
\begin{align}
\qrel{\rho_p}{\rho_n(X^n)}
\leq \tr{(\rho_p-\rho_n)\rho_n^{-1}(\rho_p-\rho_n)}
\leq 2\kappa_p\norm{\rho_n(X^n)-\rho_p}_2^2. \label{eq:KLbnd-chisq}
\end{align}
Using the matrix Bernstein bound in\eqref{eq:concmatbernstein}, we can upper bound the probability of the complementary event $\cG_n^c$ as
\begin{align}
\PP(\mathcal G_n^c)\leq 2d\,e^{-\left(\frac{n\delta_n^2}{4}\wedge\frac{3n\delta_n}{4}\right)}.\notag
\end{align}
Finally, by independence and $\norm{\varphi(X)}=1$,
\begin{align}
\EE\norm{\rho_n(X^n)-\rho_p}_2^2
\leq \left(1-\frac1n\right)^2\frac{1-\tr{\rho_p^2}}{n}
+\frac{\norm{\pi_d-\rho_p}_2^2}{n^2}\leq\frac 1n. \notag 
\end{align}
Combining these estimates with \eqref{eq:bndqrelmain} gives,  for $n>2\kappa_p$,
\begin{align}
\EE\left[\qrel{\rho_p}{\hat{\sigma}_n^{\star}(X^n)}\right]
\leq \frac{2\kappa_p+\log d}{n}
+2d\log(dn)e^{-\left(\frac{n\delta_n^2}{4}\wedge\frac{3n\delta_n}{4}\right)}. \notag 
\end{align}
For $n\leq2\kappa_p$, the crude bound $\qrel{\rho_p}{\rho_n(X^n)}\leq\log(dn)$ (due to $\rho_n(X^n)\geq I/(dn)$) together with \eqref{eq:bndqrelmain} remains valid. 

\medskip
Next, we prove \eqref{eq:devineqqrel} given \eqref{eq:asqrelbnd} holds. Under this assumption, following the same steps as in expectation bounds in Part $(ii)$ for the case $\epsilon>0$ yields almost surely that
\begin{align}
     \qrel{\rho_p}{\hat \sigma_n^{\star}(X^n)} \leq \frac{2\bar{b}_n}{n}+ d \bar{b}_n \norm{\hat{\rho}_n(X^n)-\rho_p}+\frac{\log d}{n}+\epsilon. \notag 
\end{align}
Hence, \eqref{eq:devineqqrel} follows:
\begin{align}
    \PP\left( \qrel{\rho_p}{\hat \sigma_n^{\star}(X^n)} \geq \frac{2\bar{b}_n}{n}+\frac{\log d}{n}+\epsilon+d \bar{b}_n t  \right) &\leq  \PP\left(  \norm{\hat{\rho}_n(X^n)-\rho_p} \geq t \right) \leq 2d e^{-\left(\frac{n t^2}{4} \wedge \frac{3nt}{4}\right)}.\notag 
\end{align}
Finally, let $n>2\kappa_p$ and $0\leq t\leq\delta_n$. Under the event $\norm{\hat\rho_n(X^n)-\rho_p}\leq t$, the same argument as in \eqref{eq:KLbnd-chisq} gives
\begin{align}
\qrel{\rho_p}{\rho_n(X^n)}\leq 2d\kappa_p\left(t+\frac1n\right)^2. \notag
\end{align}
Together with \eqref{eq:bndqrelmain}, this yields
\begin{align}
\PP\left( \qrel{\rho_p}{\hat \sigma_n^{\star}(X^n)} \geq 2d\kappa_p\left(t+\frac1n\right)^2+\frac{\log d}{n}  \right)
\leq \PP\left(\norm{\hat\rho_n(X^n)-\rho_p}>t\right)
\leq 2d e^{-\left(\frac{n t^2}{4} \wedge \frac{3nt}{4}\right)}, \notag
\end{align}
which proves \eqref{eq:devineqqrelmatched}.
\subsection{Proof of Proposition \ref{Prop:covnumdensop}} \label{Sec:Prop:covnumdensop-proof}
Let $\rho=U_{\rho}\Lambda_{\rho}U_{\rho}^{\dag}$ and $\sigma=U_{\sigma}\Lambda_{\sigma}U_{\sigma}^{\dag}$. 
We have
\begin{align}
    \norm{\rho-\sigma}_2^2 
&= \norm{U_{\rho}\Lambda_{\rho}U_{\rho}^{\dag}-U_{\sigma}\Lambda_{\sigma}U_{\sigma}^{\dag}}_2^2\notag \\
    &\stackrel{(a)}{\leq}   3\norm{U_{\rho}\Lambda_{\rho}U_{\rho}^{\dag} -U_{\rho}\Lambda_{\rho}U_{\sigma}^{\dag}}_2^2+3\norm{U_{\rho}\Lambda_{\rho}U_{\sigma}^{\dag}-U_{\rho}\Lambda_{\sigma}U_{\sigma}^{\dag}}_2^2+3\norm{U_{\rho}\Lambda_{\sigma}U_{\sigma}^{\dag}-U_{\sigma}\Lambda_{\sigma}U_{\sigma}^{\dag}}_2^2 \notag \\
        &\stackrel{(b)}{\leq}   3\norm{U_{\rho}\Lambda_{\rho}U_{\rho}^{\dag} -U_{\rho}\Lambda_{\rho}U_{\sigma}^{\dag}}_1^2+3\norm{U_{\rho}\Lambda_{\rho}U_{\sigma}^{\dag}-U_{\rho}\Lambda_{\sigma}U_{\sigma}^{\dag}}_2^2+3\norm{U_{\rho}\Lambda_{\sigma}U_{\sigma}^{\dag}-U_{\sigma}\Lambda_{\sigma}U_{\sigma}^{\dag}}_1^2 \notag \\
    & \stackrel{(c)}{\leq} 3 \norm{U_{\rho}}^2 \norm{\Lambda_{\rho}}_1^2\norm{U_{\rho}^{\dag}- U_{\sigma}^{\dag}}^2+3\norm{\Lambda_{\rho}- \Lambda_{\sigma}}_2^2 +3\norm{U_{\rho}-U_{\sigma}}^2 \norm{\Lambda_{\sigma}}_1^2 \norm{U_{\sigma}^{\dag}}^2 \notag \\
    &\stackrel{(d)}{=}  6 \norm{U_{\rho}- U_{\sigma}}^2 +3\norm{\Lambda_{\rho}- \Lambda_{\sigma}}_2^2, \label{eq:tracedistbnd}
\end{align}
where $(a)$ is via triangle inequality for Schatten norms and using $(a+b+c)^2 \leq 3(a^2+b^2+c^2)$ for any $a,b,c \in \RR$; $(b)$ is due to $\norm{A}_2 \leq \norm{A}_1$; $(c)$ is via an application of Cauchy-Schwarz inequality  and   unitary invariance of Schatten norms; and $(d)$ is because $\norm{\Lambda_{\rho}}_1=\norm{U}=\norm{U^{\dag}}=1$ for a unitary $U$ and density operator $\rho$.

From \eqref{eq:tracedistbnd} and the chi-squared upper bound for quantum relative entropy in \eqref{eq:chisquppbnd}, it follows that
\begin{align}
\qrel{\rho}{\sigma} 
& \leq \tr{(\rho-\sigma)^2\sigma^{-1}} \notag \\
 & \leq \norm{\sigma^{-1}} \norm{\rho-\sigma}_2^2 \notag \\
 &\leq \norm{\sigma^{-1}} \left(6 \norm{U_{\rho}- U_{\sigma}}^2+ 3\norm{\Lambda_{\rho}- \Lambda_{\sigma}}_2^2\right)\notag \\
 & = 6 \norm{U_{\rho}- U_{\sigma}}^2 \norm{\Lambda_{\sigma}^{-1}}+ 3\norm{\Lambda_{\rho}- \Lambda_{\sigma}}_2^2 \norm{\Lambda_{\sigma}^{-1}}. \label{eq:ubndqrelcov} 
\end{align}
Based on the above equation, we will construct an $\epsilon$-covering for $\cS(\HH_d)$ in terms of a covering of the unitary group $\UU_d$  and a covering of the probability simplex $\cP(\cZ)$ with $\cZ=\{1,\cdots,d\}$.

Note that $\norm{U}=1$ for all $U \in \UU_d$. Also, using the relation $U=e^{iH}$ for a Hermitian matrix, it follows that the degrees of freedom of $U$ is $d^2$, and moreover, it is sufficient to restrict to $\cH\coloneqq\{H:\norm{H} \leq \pi\}$, i.e., $\UU_d=\{U_H: U_H \coloneqq e^{iH},~H \in \cH\}$. Next, observe that since the Lipschitz constant of the map $H \mapsto e^{iH}$ is bounded by $1$, we have  $\norm{U_{H_1}-U_{H_2}} \leq \norm{H_1-H_2}$. Hence,  to obtain an upper bound on the $\epsilon$-covering number of $\UU_d$, it is sufficient to obtain it for $N(\epsilon, \cH, \norm{\cdot})$. 

Similar to volumetric arguments used to obtain covering numbers of unit ball in  Euclidean distance, 
 we have for all $0<\epsilon \leq 1$ that
  \begin{align}
      N(\epsilon, \UU_d, \norm{\cdot}) \leq  N(\epsilon, \cH, \norm{\cdot}) \stackrel{(a)}{\leq}\left(1+2\pi\epsilon^{-1}\right)^{d^2} \leq \left((1+2\pi) \epsilon^{-1}\right)^{d^2}, \notag
 \end{align} 
where  $(a)$ used \citep[Proposition 4.2.12]{Vershynin2018HighDimensionalProbability} and the fact that $\cH$ is a real $d^2$- dimensional normed space such that each element of it has operator norm bounded by $\pi$. Hence, there exists  an $\epsilon$-covering of $\UU_d$, $\cU(\epsilon)$, satisfying
\begin{align}
  |\cU(\epsilon)|  \leq \left((1+2\pi) \epsilon^{-1}\right)^{d^2}. \notag
\end{align}
Let $\gamma\in\NN$ satisfy $\gamma>d-1$. Let $\cD(\gamma)$ be the set of pmfs $q \in \cP(\cZ)$ such that $q(i)=j_i/\gamma$ for positive integers $j_i \in \NN$. As shown in \citep[Equation 2.55]{tang2022divergence},  for any $p \in \cP(\cZ)$, there exists $q \in \cD(\gamma)$ such that
\begin{align}
    \norm{p- q}_2^2 \norm{q^{-1}} \leq \sum_{i=1}^{d-1} \frac{1/\gamma^2}{1/\gamma}+\frac{(d-1)^2/\gamma^2}{1/\gamma}=\frac{d(d-1)}{\gamma}. \label{eq:KLcovchisq} 
\end{align}
Choose
$\gamma=\left\lceil 6d(d-1)/\epsilon\right\rceil$. Then the RHS of \eqref{eq:KLcovchisq} is at most $\epsilon/6$. Moreover, for $d\geq2$ and $0<\epsilon\leq1$, $\gamma\leq \frac{13}{2}\frac{d(d-1)}{\epsilon}$, and hence
\begin{align}
|\cD(\gamma)| \leq \binom{\gamma-1}{d-1} \leq \left(\frac{\gamma e}{d-1}\right)^{d-1} \leq \left(\frac{13 e d}{2\epsilon}\right)^{d-1}. \notag
\end{align}
Set
\begin{align}
  \cA=\left\{U\Lambda_qU^{\dag}: U \in \cU\big(\sqrt{\epsilon/(12 \gamma)}\big),~q \in \cD(\gamma)\right\}.  \notag
\end{align}
Then, \eqref{eq:KLcovchisq} along with the construction of $\cD(\gamma)$ and  $\cA$ implies that  for any $\rho \in \cS(\HH_d)$ with $\Lambda_{\rho}=\mathrm{diag}(p)$, there exists $q \in \cD(\gamma)$ and $\sigma \in  \cA$ such that $\Lambda_{\sigma}=\mathrm{diag}(q)$,  $\norm{\Lambda_{\rho}- \Lambda_{\sigma}}_2^2 \norm{\Lambda_{\sigma}^{-1}} \leq \epsilon/6$, and $\norm{U_{\rho}- U_{\sigma}}^2 \norm{\Lambda_{\sigma}^{-1}} \leq \epsilon/12$ (since $\norm{\Lambda_{\sigma}^{-1}} \leq \gamma$). Hence, for such a $\sigma$, we have
\begin{align}
\qrel{\rho}{\sigma} \leq  6 \norm{U_{\rho}- U_{\sigma}}^2 \norm{\Lambda_{\sigma}^{-1}}+ 3\norm{\Lambda_{\rho}- \Lambda_{\sigma}}_2^2 \norm{\Lambda_{\sigma}^{-1}}  \leq 0.5 \epsilon+ 0.5\epsilon=\epsilon.\notag 
\end{align}
Thus, $\cA$ is an $\epsilon$-covering of $\cS(\HH_d)$, and the proof is completed by noting that
\begin{align}
|\cA| \leq \left((13/2) e d\epsilon^{-1}\right)^{d-1} \left((1+2\pi)\sqrt{78}\,d \epsilon^{-1}\right)^{d^2} \leq \left(c d \epsilon^{-1}\right)^{d^2+d-1}. \notag
\end{align}
Since $(1+2\pi)\sqrt{78}<65$, the conclusion of Proposition \ref{Prop:covnumdensop} holds with a universal constant $c<65$.

\subsection{Proof of Theorem \ref{Thm:quantPyth} and Theorem \ref{Thm:Qpyth-infdim}} \label{Sec:Thm:quantPyth-proof}
Consider $\rho_0,\rho_1,\sigma\in\cS(\HH)$ such that $\mathsf D(\rho_0\|\sigma) \vee \mathsf D(\rho_1\|\sigma) <\infty$. For $t\in(0,1)$, set $\rho_t=(1-t)\rho_0+t\rho_1$.  By  Donald's identity \citep{Donald1987FreeEA} 
\begin{align}
(1-t)\mathsf D(\rho_0\|\sigma)+t\mathsf D(\rho_1\|\sigma)
=\mathsf D(\rho_t\|\sigma)+(1-t)\mathsf D(\rho_0\|\rho_t)+t\mathsf D(\rho_1\|\rho_t). \label{eq:donald-two-state}
\end{align}

First, we prove Theorem \ref{Thm:Qpyth-infdim}. By compactness and lower semicontinuity, a minimizer $\rho^{\star}\coloneqq\rho^{\star}_{\sigma,\bar{\cS}}$ exists and \eqref{eq:infdim-finite-inf} implies $\mathsf D(\rho^{\star}\|\sigma)<\infty$. Fix $\rho\in\hat{\cS}$ and define $\rho_t=(1-t)\rho^{\star}+t\rho$. Convexity of quantum relative entropy implies  $\rho_t\in\bar{\cS}$, and optimality of $\rho^{\star}$ yields
\begin{align}
\mathsf D(\rho_t\|\sigma)\geq\mathsf D(\rho^{\star}\|\sigma). \notag
\end{align}
Applying \eqref{eq:donald-two-state} to $\rho^{\star}$ and $\rho$ and using the above equation, we obtain
\begin{align}
\mathsf D(\rho\|\sigma)-\mathsf D(\rho^{\star}\|\sigma)
\geq \frac{1-t}{t}\mathsf D(\rho^{\star}\|\rho_t)+\mathsf D(\rho\|\rho_t)
\geq \mathsf D(\rho\|\rho_t). \notag
\end{align}
Since $\rho_t\to\rho^{\star}$ in trace norm as $t\downarrow0$, lower semicontinuity in the second argument gives
\begin{align}
\mathsf D(\rho\|\rho^{\star})\leq\liminf_{t\downarrow0}\mathsf D(\rho\|\rho_t). \notag
\end{align}
Together with the previous equation, this proves \eqref{eq:pythgthm-infdim} for $\rho\in\hat{\cS}$. If $\rho\notin\hat{\cS}$, the left-hand side of \eqref{eq:pythgthm-infdim} is  $\infty$, so the inequality holds trivially. In particular, for each $\rho\in\hat{\cS}$, \eqref{eq:pythgthm-infdim}  yields  $\mathsf D(\rho\|\rho^{\star})<\infty$, which implies $\rho\ll\rho^{\star}$. Therefore $\supp(\hat{\cS})\subseteq\supp(\rho^{\star})$ and the reverse inclusion follows from $\rho^{\star}\in\hat{\cS}$. Hence $\supp(\rho^{\star})=\supp(\hat{\cS})$. Finally, if $\rho_1^{\star}$ and $\rho_2^{\star}$ are minimizers, applying the Pythagorean inequality with $\rho=\rho_2^{\star}$ and reference minimizer $\rho_1^{\star}$ gives $\mathsf D(\rho_2^{\star}\|\rho_1^{\star})=0$, hence uniqueness.

Next, we prove  Theorem \ref{Thm:quantPyth}, starting with Part $(i)$. In finite dimensions, $\supp(\bar{\cS})\subseteq\supp(\sigma)$ implies $\mathsf D(\rho\|\sigma)<\infty$ for every $\rho\in\bar{\cS}$. Thus, Theorem \ref{Thm:Qpyth-infdim} yields \eqref{eq:pythgthm} and $\supp(\rho^{\star})=\supp(\bar{\cS})$. Suppose in addition that $\bar{\cS}$ is linearly closed.  For any $\rho\in\bar{\cS}$, positivity of $\rho^{\star}$ on  $\supp(\bar{\cS})$,  gives an $s>0$ such that $\eta=(1+s)\rho^{\star}-s\rho$ is a density operator. Linear closedness then implies that $\eta\in\bar{\cS}$. Then, $\rho^{\star}$ can be written as a convex combination 
\begin{align}
\rho^{\star}=\frac{s}{1+s}\rho+\frac{1}{1+s}\eta. \notag
\end{align}
 Applying Donald's identity with reference state $\sigma$ shows that
\begin{align}
 \frac{s}{1+s}\qrel{\rho}{\sigma} +\frac{1}{1+s} \qrel{\eta}{\sigma} = \qrel{\rho^{\star}}{\sigma} + \frac{s}{1+s}\qrel{\rho}{\rho^{\star}} +\frac{1}{1+s} \qrel{\eta}{\rho^{\star}}. \notag 
\end{align}
From these two equations, we obtain
\begin{align}
 \frac{s}{1+s}\left(\qrel{\rho}{\sigma}-\qrel{\rho}{\rho^{\star}}-\qrel{\rho^{\star}}{\sigma} \right) +\frac{1}{1+s} \left(\qrel{\eta}{\sigma}-\qrel{\eta}{\rho^{\star}}-\qrel{\rho^{\star}}{\sigma} \right)  =   0. \notag  
\end{align}
Since $\qrel{\rho}{\sigma} \geq \qrel{\rho}{\rho^{\star}}+\qrel{\rho^{\star}}{\sigma}$ and $\qrel{\eta}{\sigma} \geq \qrel{\eta}{\rho^{\star}}+\qrel{\rho^{\star}}{\sigma}$ holds by \eqref{eq:pythgthm-infdim}, the above equation implies that $\qrel{\rho}{\sigma} = \qrel{\rho}{\rho^{\star}}+\qrel{\rho^{\star}}{\sigma}$ and $\qrel{\eta}{\sigma} = \qrel{\eta}{\rho^{\star}}+\qrel{\rho^{\star}}{\sigma}$, thus proving equality in \eqref{eq:pythgthm}. If $\supp(\bar{\cS})\not\subseteq\supp(\sigma)$, the statement in Remark \ref{rem:suppcond} follows by restricting to $\bar{\cS}_{\sigma}=\{\rho\in\bar{\cS}:\rho\ll\sigma\}$.

\medskip

Next, we prove Part $(ii)$. We will denote $\cM(\rho_0,\cC)$ by $\cM$,  $\widebar{\cM}_{\sigma}(\rho_0,\cC)$  by $\widebar{\cM}_{\sigma}$, and $\mathsf{P}_{\widebar \cM_{\sigma}}$ by $\mathsf{P}$, for simplicity.   Note that since $\cM$ is  compact (due to being closed and bounded in $\cL(\HH_d)$) and linearly closed, so is $\widebar{\cM}_{\sigma}$.  By Part $(i)$, the optimizer $\rho^{\star}\coloneqq\rho^{\star}_{\sigma,\widebar{\cM}_{\sigma}}$ is strictly positive on  $\supp(\rho^{\star})=\supp(\widebar{\cM}_{\sigma})$, and the Pythagorean equality holds on $\widebar{\cM}_{\sigma}$. Hence, the proof will be completed if we show that 
\begin{align}
\rho^{\star}
\in \mathrm{Exp}\big(\omega_{\sigma,\mathsf{P}},\mathsf{P}\cC_{\mathsf{H}} \mathsf{P}\big), \label{eq:expfamopt}  
\end{align}
where $\omega_{\sigma,\mathsf{P}}$ is as defined in \eqref{eq:expfam-reference}. 

To prove this, let $\alpha_i \coloneqq \tr{\rho_0 L_i}$ and note  that every $\rho \in \widebar{\cM}_{\sigma}$  should satisfy
 \begin{align}
     \tr{\rho(L_i-\alpha_iI)}=  \tr{\rho\mathsf{P}(L_i-\alpha_iI)\mathsf{P}}=0, \label{eq:orthcond1}
 \end{align}
 where the first equality is due to  cyclicity of trace  and $\supp(\rho) \subseteq \supp(\widebar{\cM}_{\sigma})$.  
This is equivalent to  \begin{align}
     \tr{\rho\mathsf{P}\left(L_i^{\mathrm{r}}-\alpha_i^{\mathrm{r}}I\right)\mathsf{P}}=\tr{\rho\mathsf{P}\left(L_i^{\mathrm{i}}-\alpha_i^{\mathrm{i}}I\right)\mathsf{P}}=0, \notag
 \end{align}
 where $L_i^{\mathrm{r}}\coloneqq 0.5(L_i+L_i^{\dag})$, $L_i^{\mathrm{i}}\coloneqq -0.5 \,\mathrm{i} \,(L_i-L_i^{\dag})$, and $\alpha_i^{\mathrm{r}}$ ($\alpha_i^{\mathrm{i}}$) denotes the real (imaginary) parts of $\alpha_i$. 
 Also, since  equality in \eqref{eq:pythgthm} holds with $\cS=\widebar \cM_{\sigma}$, we have
 \begin{align}
    \tr{\rho  \,\mathsf{P}\big(\log \rho^{\star}-\log \sigma-\mathsf{D}\big(\rho^{\star}\|\sigma\big) I\big) \mathsf{P}}=0. \label{eq:orthcond2}
\end{align}
Let $\cA$ denote the real subspace spanned by $\{\mathsf{P}(L_i^{\mathrm{r}}-\alpha_i^{\mathrm{r}} I)\mathsf{P}\}_{i=1}^k \cup \{ \mathsf{P}(L_i^{\mathrm{i}}-\alpha_i^{\mathrm{i}} I)\mathsf{P}\}_{i=1}^k$ . Note that \eqref{eq:orthcond1} means that $\rho \in \widebar{\cM}_{\sigma}$ is orthogonal to $\cA$ with respect to Hilbert-Schmidt inner product. Moreover, denoting by  $\cA^{\perp}$, the orthogonal complement  of $\cA$,  density operators within $\widebar \cM_{\sigma}$ span the real vector space of self-adjoint operators $\cA^{\perp}$. This can be seen by considering an orthogonal self-adjoint basis for this space, adding a scalar multiple of $\rho^{\star}$ to its each element to obtain a set of  positive operators on $\supp(\widebar{\cM}_{\sigma})$, and normalizing each such operator by the sum of its eigenvalues, to yield the desired basis of density operators (including $\rho^{\star}$).    
Given this, \eqref{eq:orthcond2} holding for all $\rho \in \widebar{\cM}_{\sigma}$  implies that 
\begin{align}
  \log \rho^{\star}-\mathsf{P}\,\log \sigma \,\mathsf{P}-\mathsf{D}\big(\rho^{\star}\|\sigma\big) \mathsf{P} =\sum_{i=1}^k \big(\beta_i^{\mathrm{r}}\, \mathsf{P}\left(L_i^{\mathrm{r}}-\alpha_i^{\mathrm{r}}I\right)\mathsf{P}+\beta_i^{\mathrm{i}}\, \mathsf{P}\left(L_i^{\mathrm{i}}-\alpha_i^{\mathrm{i}}I\right)\mathsf{P}\big),\notag
\end{align}
for some $\bm{\beta}^{\mathrm{r}},\bm{\beta}^{\mathrm{i}} \in \RR^{k}$, where $\rho^{\star}$ is considered as an operator on $\supp(\widebar{\cM}_{\sigma})$. 
Noting that $\supp(\rho^{\star}) =\supp(\widebar \cM_{\sigma})$, we obtain
\begin{align}
\rho^{\star}_{\sigma,\widebar{\cM}_{\sigma}}= \frac 1c\,e^{\mathsf{P}\log \sigma \mathsf{P}+\sum_{i=1}^k \big(\beta_i^{\mathrm{r}}\, \mathsf{P}L_i^{\mathrm{r}}\mathsf{P}+\beta_i^{\mathrm{i}}\, \mathsf{P}L_i^{\mathrm{i}}\mathsf{P}\big)}, \quad \mbox{with} \quad c=\operatorname{Tr}\Big[e^{\mathsf{P}\log \sigma \mathsf{P}+\sum_{i=1}^k \big(\beta_i^{\mathrm{r}}\, \mathsf{P}L_i^{\mathrm{r}}\mathsf{P}+\beta_i^{\mathrm{i}}\, \mathsf{P}L_i^{\mathrm{i}}\mathsf{P}\big)}\Big],\label{eq:infoproj-exp}
\end{align}
where  $c$ is the normalization constant to ensure unit trace and $\{\beta_i^{\mathrm{r}},\beta_i^{\mathrm{i}}\}_{i=1}^k$ is such that $\sum_{i=1}^k \big(\beta_i^{\mathrm{r}}\, \mathsf{P}L_i^{\mathrm{r}}\mathsf{P}+\beta_i^{\mathrm{i}}\, \mathsf{P}L_i^{\mathrm{i}}\mathsf{P}\big)$ is self-adjoint.  This implies \eqref{eq:expfamopt}. For $\rho\in\cM\setminus\widebar{\cM}_{\sigma}$, both $\mathsf D(\rho\|\sigma)$ and $\mathsf D(\rho\|\rho^{\star})$ are infinite, and so \eqref{eq:pythgthmmix} holds trivially.

\medskip

Finally, we prove Part $(iii)$. Take $L_{ij}=\ket{e_i(\rho)}\bra{e_j(\rho)}$ for $i\neq j$. The associated mixture family is precisely the set of density operators diagonal in the eigenbasis $\cE_{\rho}$. Since $\sigma>0$, $\supp(\widebar{\cM}_{\sigma})=\HH_d$,  $\mathsf{P}=I$ and $\omega_{\sigma,P}=\sigma$. Equation \eqref{eq:infoproj-exp} shows that the off-diagonal multipliers cancel the off-diagonal entries of $\log\sigma$ in the basis $\cE_{\rho}$, leaving the diagonal operator $\sum_i\mathsf{P}_i(\rho)\log\sigma\,\mathsf{P}_i(\rho)$. Normalization then gives \eqref{eq:diag-Iproj-corrected}.
\section{Concluding Remarks}
We  studied QMLP, a quantum version of MLP, formulated via embedding of empirical probability distributions into a learned Hilbert space.  We established statistical performance guarantees for QMLP given an arbitrary embedding. For the related problem of information projection, we generalized the Pythagorean theorem to mixture families with possibly non-self-adjoint constraints, and extended the associated inequality to infinite-dimensional settings. Our performance guarantees for QMLP indicate that a good embedding should aim for  minimal embedded Hilbert space dimension and  maximal smallest eigenvalue for the embedded density operator. While this provides some insights into the choice of an efficient embedding, often it is desirable for embeddings to have additional properties such as injectivity  or approximate relative distance preserving property within specific sub-classes of probability distributions.

In the context of embeddings into an RKHS generated by a  symmetric positive definite translation-invariant kernel, a necessary and sufficient condition that the corresponding feature map is injective (or kernel is characteristic) for sub-classes of distributions was obtained by \citet{GBRSS-2006,Sriperumbudur-2008-InjectiveHS,Fukumizu-Bach-Jordan-2009}. Relaxing the assumption of translation-invariance of kernels, integral strict positive definiteness was shown to be a sufficient condition for a kernel to be characteristic by \citet{SGFSL-2010}. However, the latter condition is tailored towards the kernels being injective over the entire $\cP(\cX)$, and hence is too restrictive if the class of probability distributions $P$ of interest is a strict subset. Hence, it is worthwhile to investigate when embeddings are injective for specific sub-classes of $\cP(\cX)$ when the associated kernel is not translation invariant.

A simple criteria for injectivity of covariance embedding can be obtained as follows. Suppose $\cP \subseteq \cP(\cX)$ is a class of discrete or continuous distributions on $\RR^m$ of interest. Consider the class of pairwise differences of densities (with respect to counting or Lebesgue measure) within $\cP$, i.e., $\cP^2\coloneqq \big\{ p-p': P \neq P',~P,P' \in \cP   \big\} $. Then, it is clear that the covariance embedding $\phi: \cP \rightarrow \cL(\HH_d) $ induced by feature map $\varphi:\cX \rightarrow \HH_d$ is injective on $\cP$ if and only if the intersection of $\cP^2$ and the kernel of $\phi$ is empty. Equivalently, injectivity means  the intersection of $\cP^2$ and the kernel of the linear integral map induced by the symmetric positive definite  kernel (see Appendix \ref{Sec:RKHS}), $|K(x,y)|^2=|\innp{\varphi(x)}{\varphi(y)}|^2$,  is empty. It would be fruitful to explore the interplay between specific kernels and  $\cP$ for which they are injective from this viewpoint or determine  alternative criteria that are simpler to verify.

Another aspect pertains to relaxing the assumption of  independent and identically distributed samples to incorporate Markov or other kinds of dependence, similarly to the classic results in \citet{Atteson-1999} extending \citep{Clarke-Barron-1990} and recent ones such as \citet{Han-Jana-Wu-2023,han2024prediction}. Yet another avenue to explore pertains to reverse contraction coefficients for specific embeddings which quantify how much the quantum relative entropy in the embedded domain is preserved compared to KL divergence in the distribution space (see the recent study by \citet{belzig2025reverse} for quantum channels).  

\medskip

\section*{Acknowledgement}
S. Sreekumar is partially supported  by the CNRS-L2S-CentraleSup\'elec funding WRP630. N. Weinberger is partially supported by the Israel Science Foundation (ISF), grant no. 1782/22. 

\medskip

\appendix

\section{Classical and Quantum Transformer-based Large Language Models}\label{Sec:LLMs}
 Modern LLMs are transformer-based models  composed of an input (embedding) layer followed by  multi-head (masked self) attention units, feedforward nets, and finally an output layer which produces probabilities over a vocabulary. The input layer converts data (text, images etc.) into tokens and further represents them as vectors in a Euclidean space.  Each multi-head attention unit consists of a stack of attention units which are processed in parallel, each head performing independent projections of tokens within the context window into some internal representation space and computing  correlations among tokens, which are concatenated at its output. To describe in more detail, let $\cW$ be a finite set of tokens and $k$ be the context window. The input embedding layer stacks a sequence of tokens $w_1, \ldots,w_k$ as a $k \times d_{in}$ matrix  $Z=(z_1,\ldots,z_k)^T$, where the row vector $z_j \in \RR^{d_{in}}$ is the representation of token $w_j$ for $1 \leq j \leq k$. Each attention head mechanism involves three matrices of dimension $d_{in} \times d_{mod}$, the query matrix $M_Q$, key matrix $M_K$ and the value matrix $M_V$,  which are learned during the training phase. Using these, the output of the attention mechanism is computed as 
 \begin{align}
A=\textrm{softmax}\left(\textrm{mask}\left(\frac{QK^T}{\sqrt{d_{mod}}}\right)\right) V, \label{eq:attentmech}
 \end{align}
 where $Q=Z W_Q$, $K=ZW_K$ and $V=Z W_V$. In the above, the masking operation enforces causality (i.e. dependence of tokens only on past tokens) by setting the relevant matrix entries to a sufficiently large negative value and softmax performs the standard softmax operation (taking exponential of each entry and normalizing by row sum to produce a probability distribution). The output from the attention units (say $L$ of them) are concatenated into a matrix of size $k \times L d_{mod}$, each row of which is processed by a feedforward network. Finally, the output projection matrix $W_O$ takes as input the output of the feedforward net of size $k \times L d_{mod}$ and outputs a matrix of size $k \times |\cV|$, where the rows correspond to the prediction probabilities over the vocabulary $\cV$.

 Quantum LLMs (QLLMs) have a similar mechanism when the input layer operation is viewed as embeddings into a Hilbert space. The desired classical and quantum computations such as \eqref{eq:attentmech} are then implemented using a hybrid classical-quantum architecture involving classical computations and quantum gates (which are unitary operators on the underlying Hilbert space). These operations induce a quantum state as input to the output layer to which a POVM $\cM=\{M_v\}_{v \in \cV}$ is applied, which produces outcomes according to some probability distribution over the vocabulary $\cV$.  Denoting the set of all possible output probability distributions induced by such measurements by $\hat {\cQ}_n$, the QLLM optimization problem to find the optimal predictor is equivalent to solving \eqref{eq:mlest-emp}. 

 \section{Reproducing Kernel Hilbert Space (RKHS) and Covariance Embedding}\label{Sec:RKHS}
While reproducing kernels \citep{aronszajn-RKHS-1950} have appeared in several different contexts in learning theory and machine learning, their utility for the task of prediction has remained underexplored.    Consider a kernel function $K:\cX \times \cX \rightarrow \CC$ (or $\RR$) which satisfies the following:
\begin{itemize}
\item $K$ is a continuous and  positive definite kernel on $\cX$, i.e., 
\begin{align}
   K(x,y)=K(y,x)^{*} \quad \mbox{and} \quad  \sum_{i,j=1}^n c_i^{*}c_j K(x_i,x_j) \geq 0,
\end{align}
for every $x^n \in \cX^n$ and $c^n \in \CC^n$ (or $\RR^n$), such that $K(x,x)=1$ for all $x \in \cX$. 
\end{itemize} 
Every such $K$ defines a Hilbert space $\HH$ of functions $f:\cX \rightarrow \CC$ (or $\RR$), called an RKHS, as well as a map $\varphi:\cX \rightarrow \HH$ known as the feature map, such that for all $x,y \in \cX$,
\begin{align}
    f(x)=\innp{K(\cdot,x)}{f}, \quad \varphi(x) \coloneqq K(\cdot,x), \quad \mbox{and}\quad  K(x,y) \coloneqq \innp{K(\cdot,x)}{K(\cdot,y)}. 
\end{align}
Any function $f$ in $\HH$ can be obtained as linear combinations of kernel functions, i.e., 
\begin{align}
    f(x)=\sum_{i=1}^k \alpha_i K(x,x_i), \quad \mbox{for some} \quad \bm{\alpha} \in \CC^k,\bm{x} \in \cX^k,
\end{align}
or limits of such functions in $\HH$. Note that the above properties implies that $\norm{\varphi(x)}_{\HH}=1$ for all $x \in \cX$. We refer to \citet{BerlinetThomasAgnan2004} for further details.

The feature map induces the so-called mean embedding  of probability distributions into the associated RKHS, and has been used in a variety of applications (see e.g.\citep{muandet2017kernel}). Recently, the covariance embedding $\phi$ (see Section \ref{Sec:Emb-QMLP}) induced by \eqref{eq:covembed}  was studied in \citet{Bach-2023}, which embeds probability distributions into   the subclass of density operators $\cS(\HH)$ within $\cL(\HH)$. This can be seen by noting that $\phi$ is a positive linear map, and hence the image $\rho_p$ of each probability distribution $P$ satisfies $\rho_p \geq 0$. Moreover,  $\tr{\rho_p}=1$. Denoting by  $\{e_i\}_{i=1}^{\infty}$  an orthonormal basis of $\HH$, the latter follows as 
\begin{align}
    \tr{\rho_p}\coloneqq \sum_{i=1}^{\infty} \innp{e_i}{\rho_p e_i}&= \int_{\cX} \sum_{i=1}^{\infty} \bra{e_i}\ket{\varphi(x)} \bra{\varphi(x)} \ket{e_i}p(x) d \mu(x) \notag \\
    &= \int_{\cX} \sum_{i=1}^{\infty} \innp{\varphi(x)}{e_i} \innp{e_i}{\varphi(x)}p(x) d \mu(x) \notag \\
    &= \int_{\cX} \norm{\varphi(x)}_{\HH}^2 p(x) d \mu(x) =1, \label{eq:traceone}
\end{align}
where the last equality is because $\norm{\varphi(x)}_{\HH}=1$ for every $x \in \cX$ and $\int p d \mu=1$. Density operators are trace-class and hence lies within the class of Hilbert-Schmidt operators equipped with the Hilbert-Schmidt inner-product:
\begin{align}
  \innp{\rho}{\sigma} \coloneqq \tr{\rho^*\sigma}\coloneqq \sum_{i=1}^{\infty} \innp{\rho e_i}{\sigma e_i},\forall \rho,\sigma \in \cL(\HH),  \notag
\end{align}
where  $\rho^*$ denotes the adjoint of $\rho$.
\bibliographystyle{abbrvnat}
\bibliography{ref,ref-quant,quantum_prediction}

\begin{thebibliography}{69}
\providecommand{\natexlab}[1]{#1}
\providecommand{\url}[1]{\texttt{#1}}
\expandafter\ifx\csname urlstyle\endcsname\relax
  \providecommand{\doi}[1]{doi: #1}\else
  \providecommand{\doi}{doi: \begingroup \urlstyle{rm}\Url}\fi

\bibitem[Amari and Nagaoka(2000)]{Amari-Nagaoka-2000}
S.~Amari and H.~Nagaoka.
\newblock \emph{Methods of Information Geometry}.
\newblock American Mathematical Society, Oxford University Press, 2000.

\bibitem[Amin et~al.(2018)Amin, Andriyash, Rolfe, Kulchytskyy, and Melko]{Amin-QBM-2018}
M.~H. Amin, E.~Andriyash, J.~Rolfe, B.~Kulchytskyy, and R.~Melko.
\newblock Quantum {B}oltzmann machine.
\newblock \emph{Phys. Rev. X}, 8:\penalty0 021050, May 2018.
\newblock \doi{10.1103/PhysRevX.8.021050}.

\bibitem[Aronszajn(1950)]{aronszajn-RKHS-1950}
N.~Aronszajn.
\newblock Theory of reproducing kernels.
\newblock \emph{Transactions of the American Mathematical Society}, 68\penalty0 (3):\penalty0 337--404, 1950.

\bibitem[Atteson(1999)]{Atteson-1999}
K.~Atteson.
\newblock {The asymptotic redundancy of Bayes rules for Markov chains}.
\newblock \emph{IEEE Transactions on Information Theory}, 45\penalty0 (6):\penalty0 2104--2109, 1999.
\newblock \doi{10.1109/18.782149}.

\bibitem[Bach(2023)]{Bach-2023}
F.~Bach.
\newblock Information theory with kernel methods.
\newblock \emph{IEEE Transactions on Information Theory}, 69\penalty0 (2):\penalty0 752--775, 2023.
\newblock \doi{10.1109/TIT.2022.3211077}.

\bibitem[Basile and Tamburini(2017)]{basile-tamburini-QLM}
I.~Basile and F.~Tamburini.
\newblock Towards quantum language models.
\newblock In M.~Palmer, R.~Hwa, and S.~Riedel, editors, \emph{Proceedings of the 2017 Conference on Empirical Methods in Natural Language Processing}, pages 1840--1849, Copenhagen, Denmark, Sept. 2017. Association for Computational Linguistics.
\newblock \doi{10.18653/v1/D17-1196}.

\bibitem[Belzig et~al.(2025)Belzig, Gao, Smith, and Wu]{belzig2025reverse}
P.~Belzig, L.~Gao, G.~Smith, and P.~Wu.
\newblock Reverse-type data processing inequality.
\newblock \emph{Communications in Mathematical Physics}, 406\penalty0 (12):\penalty0 295, 2025.

\bibitem[Benedetti et~al.(2017)Benedetti, Realpe-G\'omez, Biswas, and Perdomo-Ortiz]{Benedetti-2017-QBMs}
M.~Benedetti, J.~Realpe-G\'omez, R.~Biswas, and A.~Perdomo-Ortiz.
\newblock Quantum-assisted learning of hardware-embedded probabilistic graphical models.
\newblock \emph{Phys. Rev. X}, 7:\penalty0 041052, Nov 2017.
\newblock \doi{10.1103/PhysRevX.7.041052}.

\bibitem[Benedetti et~al.(2019)Benedetti, Garcia-Pintos, Perdomo, Leyton-Ortega, Nam, and Perdomo-Ortiz]{Benedetti2019generative}
M.~Benedetti, D.~Garcia-Pintos, O.~Perdomo, V.~Leyton-Ortega, Y.~Nam, and A.~Perdomo-Ortiz.
\newblock A generative modeling approach for benchmarking and training shallow quantum circuits.
\newblock \emph{npj Quantum Information}, 5\penalty0 (1):\penalty0 45, 2019.
\newblock \doi{10.1038/s41534-019-0157-8}.

\bibitem[Berlinet and Thomas-Agnan(2004)]{BerlinetThomasAgnan2004}
A.~Berlinet and C.~Thomas-Agnan.
\newblock \emph{Reproducing Kernel {H}ilbert Spaces in Probability and Statistics}.
\newblock Kluwer Academic Publishers, Boston, MA, USA, 2004.
\newblock ISBN 978-1-4020-7679-4.
\newblock \doi{10.1007/978-1-4419-9096-9}.

\bibitem[Berta et~al.(2015)Berta, Fawzi, and Tomamichel]{Berta2015OnEntropies}
M.~Berta, O.~Fawzi, and M.~Tomamichel.
\newblock On variational expressions for quantum relative entropies.
\newblock \emph{Letters in Mathematical Physics}, 107\penalty0 (12):\penalty0 2239--2265, Dec. 2015.
\newblock \doi{10.1007/s11005-017-0990-7}.

\bibitem[Cesa-Bianchi and Lugosi(2006)]{cesa2006prediction}
N.~Cesa-Bianchi and G.~Lugosi.
\newblock \emph{Prediction, learning, and games}.
\newblock Cambridge university press, 2006.

\bibitem[Clarke and Barron(1990)]{Clarke-Barron-1990}
B.~Clarke and A.~Barron.
\newblock {Information-theoretic asymptotics of Bayes methods}.
\newblock \emph{IEEE Transactions on Information Theory}, 36\penalty0 (3):\penalty0 453--471, 1990.
\newblock \doi{10.1109/18.54897}.

\bibitem[Csisz{\'a}r and Matus(2003)]{csiszar2003information}
I.~Csisz{\'a}r and F.~Matus.
\newblock Information projections revisited.
\newblock \emph{IEEE Transactions on Information Theory}, 49\penalty0 (6):\penalty0 1474--1490, 2003.

\bibitem[Csisz{\'a}r and Shields(2004)]{Csiszar-Shields-2004}
I.~Csisz{\'a}r and P.~C. Shields.
\newblock Information theory and statistics: A tutorial.
\newblock \emph{Foundations and Trends{\textregistered} in Communications and Information Theory}, 1\penalty0 (4):\penalty0 417--528, 2004.

\bibitem[Donald(1986)]{Donald1986}
M.~J. Donald.
\newblock On the relative entropy.
\newblock \emph{Communications in Mathematical Physics}, 105\penalty0 (1):\penalty0 13--34, 1986.
\newblock ISSN 1432-0916.
\newblock \doi{10.1007/BF01212339}.

\bibitem[Donald(1987)]{Donald1987FreeEA}
M.~J. Donald.
\newblock Free energy and the relative entropy.
\newblock \emph{Journal of Statistical Physics}, 49:\penalty0 81--87, 1987.

\bibitem[Fawzi and Saunderson(2023)]{Fawzi-Saunderson-2023}
H.~Fawzi and J.~Saunderson.
\newblock Optimal self-concordant barriers for quantum relative entropies.
\newblock \emph{SIAM Journal on Optimization}, 33\penalty0 (4):\penalty0 2858--2884, 2023.
\newblock \doi{10.1137/22M1500216}.

\bibitem[Fawzi et~al.(2018)Fawzi, Saunderson, and Parrilo]{Fawzi2018MatrixLog}
H.~Fawzi, J.~Saunderson, and P.~A. Parrilo.
\newblock Semidefinite approximations of the matrix logarithm.
\newblock \emph{Foundations of Computational Mathematics}, 19:\penalty0 259--296, 2018.
\newblock \doi{10.1007/s10208-018-9385-0}.

\bibitem[Fukumizu et~al.(2004)Fukumizu, Bach, and Jordan]{FukumizuBachJordan2004}
K.~Fukumizu, F.~R. Bach, and M.~I. Jordan.
\newblock Dimensionality reduction for supervised learning with reproducing kernel {H}ilbert spaces.
\newblock \emph{Journal of Machine Learning Research}, 5:\penalty0 73--99, 2004.

\bibitem[Fukumizu et~al.(2009)Fukumizu, Bach, and Jordan]{Fukumizu-Bach-Jordan-2009}
K.~Fukumizu, F.~R. Bach, and M.~I. Jordan.
\newblock Kernel dimension reduction in regression.
\newblock \emph{The Annals of Statistics}, 37\penalty0 (4):\penalty0 1871--1905, 2009.

\bibitem[Gao et~al.(2024)Gao, Ji, and Wei]{GaoJiWei2024}
M.~Gao, Z.~Ji, and F.~Wei.
\newblock Quantum maximum entropy inference and hamiltonian learning.
\newblock \emph{arXiv:2407.11473}, 2024.

\bibitem[Golowich et~al.(2020)Golowich, Rakhlin, and Shamir]{golowich2020size}
N.~Golowich, A.~Rakhlin, and O.~Shamir.
\newblock Size-independent sample complexity of neural networks.
\newblock \emph{Information and Inference: A Journal of the IMA}, 9\penalty0 (2):\penalty0 473--504, 2020.

\bibitem[Gretton et~al.(2006)Gretton, Borgwardt, Rasch, Sch\"{o}lkopf, and Smola]{GBRSS-2006}
A.~Gretton, K.~M. Borgwardt, M.~Rasch, B.~Sch\"{o}lkopf, and A.~J. Smola.
\newblock A kernel method for the two-sample-problem.
\newblock In \emph{Proceedings of the 20th International Conference on Neural Information Processing Systems}, NIPS'06, page 513–520, Cambridge, MA, USA, 2006. MIT Press.

\bibitem[Han et~al.(2023)Han, Jana, and Wu]{Han-Jana-Wu-2023}
Y.~Han, S.~Jana, and Y.~Wu.
\newblock Optimal prediction of markov chains with and without spectral gap.
\newblock \emph{IEEE Transactions on Information Theory}, 69\penalty0 (6):\penalty0 3920--3959, 2023.
\newblock \doi{10.1109/TIT.2023.3239508}.

\bibitem[Han et~al.(2024)Han, Jiang, and Wu]{han2024prediction}
Y.~Han, T.~Jiang, and Y.~Wu.
\newblock Prediction from compression for models with infinite memory, with applications to hidden markov and renewal processes.
\newblock In S.~Agrawal and A.~Roth, editors, \emph{Proceedings of the 37th Annual Conference on Learning Theory}, volume 247 of \emph{Proceedings of Machine Learning Research}, pages 2270--2307. PMLR, 2024.

\bibitem[Haussler and Opper(1997)]{haussler1997mutual}
D.~Haussler and M.~Opper.
\newblock Mutual information, metric entropy and cumulative relative entropy risk.
\newblock \emph{The Annals of Statistics}, 25\penalty0 (6):\penalty0 2451--2492, 1997.

\bibitem[Hayashi(2002)]{Hayashi_2002}
M.~Hayashi.
\newblock Optimal sequence of quantum measurements in the sense of {Stein's} lemma in quantum hypothesis testing.
\newblock \emph{Journal of Physics A: Mathematical and General}, 35\penalty0 (50):\penalty0 10759, Dec. 2002.

\bibitem[Hayashi(2016)]{Hayashi-book-2016}
M.~Hayashi.
\newblock \emph{Quantum Information Theory: Mathematical Foundation}.
\newblock Springer Berlin, Heidelberg, 2016.

\bibitem[Hayashi and Ito(2025)]{Hayashi-Ito-2025}
M.~Hayashi and Y.~Ito.
\newblock Entanglement measures for detectability.
\newblock \emph{IEEE Transactions on Information Theory}, 71\penalty0 (6):\penalty0 4385--4405, 2025.
\newblock \doi{10.1109/TIT.2025.3557056}.

\bibitem[He et~al.(2024)He, Saunderson, and Fawzi]{HeSaundersonFawzi2024}
K.~He, J.~Saunderson, and H.~Fawzi.
\newblock A bregman proximal perspective on classical and quantum blahut--arimoto algorithms.
\newblock \emph{IEEE Transactions on Information Theory}, 70\penalty0 (8):\penalty0 5710--5730, Aug. 2024.
\newblock \doi{10.1109/TIT.2024.3412129}.

\bibitem[He et~al.(2025)He, Saunderson, and Fawzi]{he2025qicsquantuminformationconic}
K.~He, J.~Saunderson, and H.~Fawzi.
\newblock {QICS: Quantum Information Conic Solver}, 2025.

\bibitem[He et~al.(2026)He, Saunderson, and Fawzi]{HeSaundersonFawzi2026}
K.~He, J.~Saunderson, and H.~Fawzi.
\newblock Interior point methods for structured quantum relative entropy optimization problems.
\newblock \emph{Mathematical Programming Computation}, 18:\penalty0 481--534, 2026.
\newblock \doi{10.1007/s12532-025-00296-w}.

\bibitem[Houska and Chachuat(2017)]{Houska-Chachuat-2017}
B.~Houska and B.~Chachuat.
\newblock Global optimization in {H}ilbert space.
\newblock \emph{Mathematical Programming}, 173, 12 2017.
\newblock \doi{10.1007/s10107-017-1215-7}.

\bibitem[Hoyos-Osorio and Sanchez-Giraldo(2024)]{Hoyos-2024}
J.~K. Hoyos-Osorio and L.~G. Sanchez-Giraldo.
\newblock The representation {Jensen-Shannon} divergence, 2024.

\bibitem[Jen\v{c}ov\'{a}(2005)]{Jencova-2005}
A.~Jen\v{c}ov\'{a}.
\newblock {Quantum information geometry and noncommutative $L_p$-spaces}.
\newblock \emph{Infinite Dimensional Analysis, Quantum Probability and Related Topics}, 08\penalty0 (02):\penalty0 215--233, 2005.
\newblock \doi{10.1142/S0219025705001949}.

\bibitem[Kachaiev and Recanatesi(2024)]{kachaiev2024learning}
O.~Kachaiev and S.~Recanatesi.
\newblock Learning to embed distributions via maximum kernel entropy.
\newblock \emph{Advances in Neural Information Processing Systems}, 37:\penalty0 44710--44734, 2024.

\bibitem[Karimi and Tun{\c c}el(2025)]{Karimi-Tuncel-2025}
M.~Karimi and L.~Tun{\c c}el.
\newblock Efficient implementation of interior-point methods for quantum relative entropy.
\newblock \emph{INFORMS J. on Computing}, 37\penalty0 (1):\penalty0 3–21, Jan. 2025.
\newblock ISSN 1526-5528.
\newblock \doi{10.1287/ijoc.2024.0570}.

\bibitem[Karimi and Tunçel(2023)]{Karimi2023DomainDrivenS}
M.~Karimi and L.~Tunçel.
\newblock Domain-driven solver (dds) version 2.1: a matlab-based software package for convex optimization problems in domain-driven form.
\newblock \emph{Mathematical Programming Computation}, 16:\penalty0 37 -- 92, 2023.

\bibitem[Kieferov\'a and Wiebe(2017)]{Kieferova-Wiebe-2017}
M.~Kieferov\'a and N.~Wiebe.
\newblock Tomography and generative training with quantum {B}oltzmann machines.
\newblock \emph{Phys. Rev. A}, 96:\penalty0 062327, Dec 2017.
\newblock \doi{10.1103/PhysRevA.96.062327}.

\bibitem[Ko{\ss}mann and Schwonnek(2026)]{KossmannSchwonnek2026}
G.~Ko{\ss}mann and R.~Schwonnek.
\newblock Optimising the relative entropy under semidefinite constraints.
\newblock \emph{npj Quantum Information}, 12\penalty0 (1):\penalty0 23, 2026.
\newblock \doi{10.1038/s41534-026-01184-4}.

\bibitem[Krichevsky and Trofimov(1981)]{Krichevsk1981performance}
R.~Krichevsky and V.~Trofimov.
\newblock The performance of universal encoding.
\newblock \emph{IEEE Transactions on Information Theory}, 27\penalty0 (2):\penalty0 199--207, 1981.
\newblock \doi{10.1109/TIT.1981.1056331}.

\bibitem[Liu and Wang(2018)]{Liu-Wang-2018}
J.-G. Liu and L.~Wang.
\newblock Differentiable learning of quantum circuit born machines.
\newblock \emph{Phys. Rev. A}, 98:\penalty0 062324, Dec 2018.
\newblock \doi{10.1103/PhysRevA.98.062324}.

\bibitem[Merhav and Feder(1998)]{merhav2002universal}
N.~Merhav and M.~Feder.
\newblock Universal prediction.
\newblock \emph{IEEE Transactions on Information Theory}, 44\penalty0 (6):\penalty0 2124--2147, 1998.

\bibitem[Muandet et~al.(2017)Muandet, Fukumizu, Sriperumbudur, and Sch{\"o}lkopf]{muandet2017kernel}
K.~Muandet, K.~Fukumizu, B.~Sriperumbudur, and B.~Sch{\"o}lkopf.
\newblock Kernel mean embedding of distributions: A review and beyond.
\newblock \emph{Foundations and Trends in Machine Learning}, 10\penalty0 (1-2):\penalty0 1--141, 2017.
\newblock \doi{10.1561/2200000060}.

\bibitem[Nielsen and Chuang(2010)]{Nielsen_Chuang_2010}
M.~A. Nielsen and I.~L. Chuang.
\newblock \emph{Quantum Computation and Quantum Information: 10th Anniversary Edition}.
\newblock Cambridge University Press, 2010.

\bibitem[Petz(1988)]{Petz1988}
D.~Petz.
\newblock A variational expression for the relative entropy.
\newblock \emph{Communications in Mathematical Physics}, 114\penalty0 (2):\penalty0 345--349, 1988.
\newblock ISSN 1432-0916.

\bibitem[Petz(2007)]{petz2007quantum}
D.~Petz.
\newblock \emph{Quantum Information Theory and Quantum Statistics}.
\newblock Springer Science \& Business Media, 2007.
\newblock \doi{10.1007/978-3-540-74636-2}.

\bibitem[Piani(2009)]{P09}
M.~Piani.
\newblock Relative entropy of entanglement and restricted measurements.
\newblock \emph{Physical Review Letters}, 103\penalty0 (16):\penalty0 160504, Oct. 2009.
\newblock \doi{10.1103/PhysRevLett.103.160504}.

\bibitem[Polyanskiy and Wu(2025)]{polyanskiy-wu-ITbook}
Y.~Polyanskiy and Y.~Wu.
\newblock \emph{Information Theory: {F}rom Coding to Learning}.
\newblock Cambridge University Press, 2025.

\bibitem[Rissanen(1983)]{rissanen2003universal}
J.~Rissanen.
\newblock Universal coding, information, prediction, and estimation.
\newblock \emph{IEEE Transactions on Information theory}, 30\penalty0 (4):\penalty0 629--636, 1983.

\bibitem[Santoro and Panaretos(2025)]{SantoroPanaretos2025}
L.~V. Santoro and V.~M. Panaretos.
\newblock Likelihood ratio tests by kernel {G}aussian embedding.
\newblock arXiv preprint, Aug 2025.

\bibitem[Smola et~al.(2007)Smola, Gretton, Song, and Sch{\"o}lkopf]{Smola-2007}
A.~Smola, A.~Gretton, L.~Song, and B.~Sch{\"o}lkopf.
\newblock A {H}ilbert space embedding for distributions.
\newblock In M.~Hutter, R.~A. Servedio, and E.~Takimoto, editors, \emph{Algorithmic Learning Theory}, pages 13--31, Berlin, Heidelberg, 2007. Springer Berlin Heidelberg.

\bibitem[Song et~al.(2010)Song, Gretton, and Guestrin]{Song-Gretton-Guestrin-2010}
L.~Song, A.~Gretton, and C.~Guestrin.
\newblock Nonparametric tree graphical models.
\newblock In Y.~W. Teh and M.~Titterington, editors, \emph{Proceedings of the Thirteenth International Conference on Artificial Intelligence and Statistics}, volume~9 of \emph{Proceedings of Machine Learning Research}, pages 765--772, Chia Laguna Resort, Sardinia, Italy, 13--15 May 2010. PMLR.

\bibitem[Sreekumar and Berta(2025)]{SB-IT-2025}
S.~Sreekumar and M.~Berta.
\newblock Limit distribution theory for quantum divergences.
\newblock \emph{IEEE Transactions on Information Theory}, 71\penalty0 (1):\penalty0 459--484, 2025.

\bibitem[Sreekumar et~al.(2026)Sreekumar, Goldfeld, and Wilde]{SGW-2026}
S.~Sreekumar, Z.~Goldfeld, and M.~M. Wilde.
\newblock Performance {G}uarantees for {Q}uantum {N}eural {E}stimation of {E}ntropies.
\newblock \emph{{Quantum}}, 10:\penalty0 2113, May 2026.
\newblock ISSN 2521-327X.
\newblock \doi{10.22331/q-2026-05-21-2113}.

\bibitem[Sriperumbudur et~al.(2008)Sriperumbudur, Gretton, Fukumizu, Lanckriet, and Scholkopf]{Sriperumbudur-2008-InjectiveHS}
B.~K. Sriperumbudur, A.~Gretton, K.~Fukumizu, G.~R.~G. Lanckriet, and B.~Scholkopf.
\newblock Injective {H}ilbert space embeddings of probability measures.
\newblock In \emph{Annual Conference Computational Learning Theory}, 2008.

\bibitem[Sriperumbudur et~al.(2010)Sriperumbudur, Gretton, Fukumizu, Sch\"{o}lkopf, and Lanckriet]{SGFSL-2010}
B.~K. Sriperumbudur, A.~Gretton, K.~Fukumizu, B.~Sch\"{o}lkopf, and G.~R. Lanckriet.
\newblock {H}ilbert space embeddings and metrics on probability measures.
\newblock \emph{J. Mach. Learn. Res.}, 11:\penalty0 1517–1561, Aug. 2010.
\newblock ISSN 1532-4435.

\bibitem[Tang(2022)]{tang2022divergence}
J.~Tang.
\newblock \emph{Divergence Covering}.
\newblock Ph.d. thesis, Massachusetts Institute of Technology, Cambridge, MA, USA, 2022.

\bibitem[Temme et~al.(2010)Temme, Kastoryano, Ruskai, Wolf, and Verstraete]{Temme-2010}
K.~Temme, M.~J. Kastoryano, M.~B. Ruskai, M.~M. Wolf, and F.~Verstraete.
\newblock {The $\chi ^2$-divergence and mixing times of quantum Markov processes}.
\newblock \emph{J. Math. Phys.}, 51\penalty0 (12):\penalty0 122201, 12 2010.

\bibitem[Thompson(1963)]{thompson_1963}
A.~C. Thompson.
\newblock On certain contraction mappings in a partially ordered vector space.
\newblock \emph{Proceedings of the American Mathematical Society}, 14:\penalty0 438--443, 1963.
\newblock ISSN 0002-9939, 1088-6826.
\newblock \doi{10.1090/S0002-9939-1963-0149237-7}.

\bibitem[Umegaki(1962)]{U62}
H.~Umegaki.
\newblock Conditional expectations in an operator algebra {IV} (entropy and information).
\newblock \emph{Kodai Mathematical Seminar Reports}, 14\penalty0 (2):\penalty0 59--85, 1962.

\bibitem[van Erven and Harremo{\"e}s(2014)]{vanEvren_Reyni_Div2014}
T.~van Erven and P.~Harremo{\"e}s.
\newblock {R}{\'e}nyi divergence and {Kullback-Leibler} divergence.
\newblock \emph{IEEE Transactions on Information Theory}, 60\penalty0 (7):\penalty0 3797--3820, Jul. 2014.

\bibitem[Vershynin(2018)]{Vershynin2018HighDimensionalProbability}
R.~Vershynin.
\newblock \emph{High-Dimensional Probability: An Introduction with Applications in Data Science}.
\newblock Cambridge University Press, Cambridge, 2018.

\bibitem[Wilde(2017)]{wilde2017quantum}
M.~M. Wilde.
\newblock \emph{Quantum Information Theory}.
\newblock Cambridge University Press, second edition, 2017.
\newblock \doi{10.1017/9781316809976.001}.

\bibitem[Wilde(2025)]{Wilde2025fundamentals}
M.~M. Wilde.
\newblock Fundamentals of quantum {B}oltzmann machine learning with visible and hidden units.
\newblock \emph{arXiv: 2512.19819}, 2025.

\bibitem[Yang and Barron(1999)]{yang1999information}
Y.~Yang and A.~Barron.
\newblock Information-theoretic determination of minimax rates of convergence.
\newblock \emph{Annals of Statistics}, pages 1564--1599, 1999.

\bibitem[You et~al.(2022)You, Cheng, and Li]{you2022minimizing}
J.-K. You, H.-C. Cheng, and Y.-H. Li.
\newblock Minimizing quantum rényi divergences via mirror descent with polyak step size.
\newblock In \emph{2022 IEEE International Symposium on Information Theory (ISIT)}, pages 252--257, 2022.
\newblock \doi{10.1109/ISIT5834325.2022.9834648}.

\bibitem[Zinchenko et~al.(2010)Zinchenko, Friedland, and Gour]{Zinchenko-2010}
Y.~Zinchenko, S.~Friedland, and G.~Gour.
\newblock Numerical estimation of the relative entropy of entanglement.
\newblock \emph{Phys. Rev. A}, 82:\penalty0 052336, Nov 2010.
\newblock \doi{10.1103/PhysRevA.82.052336}.

\end{thebibliography}
\end{document}